\documentclass[a4paper,12pt]{article}
\usepackage{epsfig}
\usepackage[dvips,usenames]{color}
\usepackage{graphicx}
\usepackage{color}
\usepackage{colortbl}

\newlength{\dinwidth}
\newlength{\dinmargin}
\newcommand{\spur}[1]{\not\! #1 \,}

\begin{document}
\title{ The rare decays $B\to K^{(*)}\bar{K}^{(*)}$ and
  R-parity violating  supersymmetry}
\author{Rumin Wang$^{1,2}$\thanks{E-mail address:
ruminwang@henannu.edu.cn }, G. R. Lu$^{1,2}$, En-Ke Wang$^{1}$, and
Ya-Dong Yang$^{2}$\thanks{E-mail address: yangyd@henannu.edu.cn }
 \\
 {\footnotesize {$^1$ \it Institute of Particle Physics,
 Huazhong Normal University,  Wuhan, Hubei 430070, P.R.China }}
  \\
{\footnotesize {$^2$ \it Department of Physics, Henan Normal
University, XinXiang, Henan 453007, P.R.China
 }}
  }
\maketitle
\begin{abstract}
We study the branching ratios, the direct $CP$ asymmetries in $B\to
K^{(*)}\bar{K}^{(*)}$ decays and the polarization fractions of $B\to
K^{*}\bar{K}^{*}$ decays by employing the QCD factorization in the
minimal supersymmetric standard model with R-parity violation. We
derive the new upper bounds on the relevant R-parity violating
couplings from the latest experimental data of $B\to
K^{(*)}\bar{K}^{(*)}$, and some of these constraints  are stronger
than the existing bounds. Using the constrained parameter spaces, we
predict the R-parity violating effects on the other quantities in
$B\to K^{(*)}\bar{K}^{(*)}$ decays which have not been measured yet.
We find that the R-parity violating effects on the branching ratios
and the direct $CP$ asymmetries could be large, nevertheless their
effects on the longitudinal polarizations of $B\to K^{*}\bar{K}^{*}$
decays are small. Near future experiments can test these predictions
and shrink the parameter spaces.
\end{abstract}

\vspace{1.5cm} \noindent {\bf PACS Numbers:  12.60.Jv,
  12.15.Mm, 12.38.Bx, 13.25.Hw}

\newpage
\section{Introduction}

The study of exclusive hadronic $B$-meson decays can provide not
only an interesting avenue to understand the $CP$ violation and
flavor mixing of the quark sector in the standard model (SM), but
also powerful means to probe different new physics (NP) scenarios
beyond the SM.  Recent experimental measurements have shown that
some $B$ decays to two light mesons deviated from the SM
expectations, for example, the $\pi\pi,\pi K$ puzzle \cite{buras}
and the polarization puzzle in $B\rightarrow VV$ decays
\cite{VVpuzzle}. Although these measurements represent quite a
challenge for theory, the SM is in no way ruled out yet since there
are many theoretical uncertainties in low energy QCD. However, it
will be under considerable strain if the experimental data persist
for a long time.

Among those NP models that survived electroweak (EW) data, one of
the most respectable options is the R-parity violating (RPV)
supersymmetry (SUSY). The possible appearance of the RPV couplings
\cite{SUSY}, which will violate the lepton and baryon number
conservation, has gained full attention in searching for SUSY
\cite{report,allanach}. The effect of  the RPV SUSY on $B$ decays
have been extensively investigated previously in the literatures
\cite{RPVstudy,hexg}, and it has been proposed as a possible
resolution to the polarization puzzle and the $\pi\pi,\pi K$ puzzle
\cite{ourVVPP}. The pure penguin $B \to K^{(*)}\bar{K}^{(*)}$ decays
are closely related with the puzzles  which are inconsistent with
the SM predictions, and therefore are very important for
understanding the dynamics of nonleptonic two-body $B$ decays, which
have been studied in Refs. \cite{KKstudy}. If the RPV SUSY is the
right model to resolve these puzzles,  the same type of NP will
affect  $B \to K^{(*)}\bar{K}^{(*)}$ decays. In this work, we shall
study the RPV SUSY effects in the $B \to K^{(*)}\bar{K}^{(*)}$
decays by using the QCD factorization (QCDF) approach \cite{BBNS}
for hadronic dynamics. The $B \to K^{(*)}\bar{K}^{(*)}$ decays are
all induced at the quark level by $b\to ds\bar{s}$ process, and they
involve the same set of  RPV coupling constants. Using the latest
experimental data and the theoretical parameters, we obtain the new
upper limits on the relevant RPV couplings. Then we use the
constrained regions of parameters to examine the RPV effects on
observations  in the $B \to K^{(*)}\bar{K}^{(*)}$ decays which have
not been measured yet.

The paper is arranged as follows.  In Sec.2, we calculate the $CP$
averaged branching ratios, the direct $CP$ asymmetries of $B \to
K^{(*)}\bar{K}^{(*)}$ and the polarization fractions in $B \to
K^{*}\bar{K}^{*}$ decays, taking  account of  the RPV effects with
the QCDF approach. In Sec.3, we tabulate the theoretical inputs in
our numerical analysis. Section 4  deals with the numerical results.
We display the constrained
 parameter spaces which satisfy all the experimental data,  and then we use the
 constrained  parameter spaces to predict the RPV effects on  the
 other observable
 quantities, which have not been measured yet in
 $B \to K^{(*)}\bar{K}^{(*)}$ system.
 Section 5 contains our summary and conclusion.

\section{The theoretical frame for $B\to K^{(*)}\bar{K}^{(*)}$decays }

\subsection{ The decay amplitudes  in the SM }
  In the SM, the low energy effective Hamiltonian for
  the $\Delta B=1$ transition at the scale $\mu$ is given by \cite{coeff}
 \begin{eqnarray}
 \mathcal{H}^{SM}_{eff}&=&\frac{G_F}{\sqrt{2}}\sum_{p=u, c}
 \lambda_p \Biggl\{C_1Q_1^p+C_2Q_2^p
 +\sum_{i=3}^{10}\Big[C_iQ_i+C_{7\gamma}Q_{7\gamma}
 +C_{8g}Q_{8g}\Big] \Biggl\}+ h.c.,
 \label{HeffSM}
 \end{eqnarray}
here  $\lambda_p=V_{pb}V_{pq}^* $ for $b \to q$ transition $(p\in
\{u,c\},q\in \{d,s\})$ and the detailed definition of the operator
base can be found in \cite{coeff}.

Using the weak effective Hamiltonian given by Eq.(\ref{HeffSM}), we
can now write the decay amplitudes for the general two-body hadronic
 $B\to M_1M_2$ decays as
\begin{eqnarray}
  \mathcal{A}^{SM}(B\to M_1M_2)&=&\left< M_1M_2|
  {\cal H}^{SM}_{eff}|B \right> \nonumber\\
  &=&\frac{G_F}{\sqrt{2}}\sum_p \sum_i \lambda_p
  C_i(\mu)\left<M_1M_2|Q_i(\mu)|B\right>.
  \end{eqnarray}
The essential theoretical difficulty for obtaining the decay
amplitude arises  from the  evaluation of hadronic matrix elements
$\langle M_1M_2|Q_i(\mu)|B\rangle$. There are at least three
approaches with different considerations to tackle the said
difficulty: the naive factorization (NF) \cite{NF1,NF2}, the
perturbative QCD \cite{PQCD}, and the QCDF \cite{BBNS}. The QCDF
 developed by Beneke, Buchalla, Neubert  and Sachrajda is
a  powerful framework for studying charmless $B$ decays.   We will
employ the  QCDF approach in this paper.

\begin{figure}[t]
\begin{center}
\includegraphics[scale=0.8]{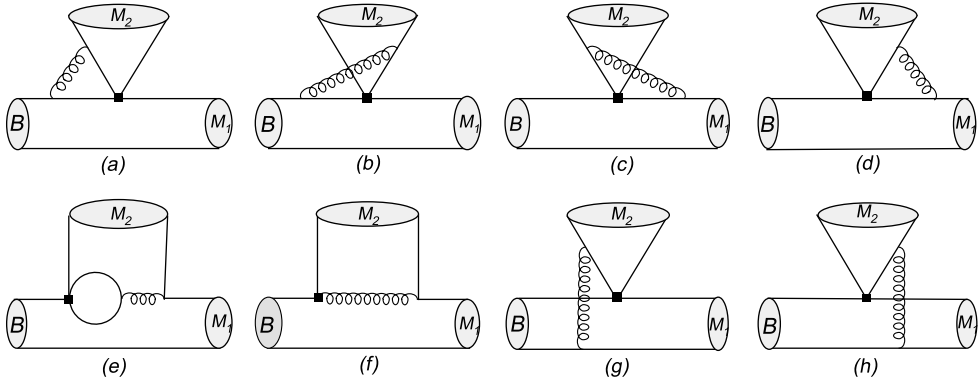}
\end{center}
\vspace{-0.6cm}
 \caption{\small The next to leading order nonfactorizable
  contributions to the coefficients $a^p_i$.}
 \label{NTL}
\end{figure}

\begin{figure}[t]
\begin{center}
\includegraphics[scale=0.8]{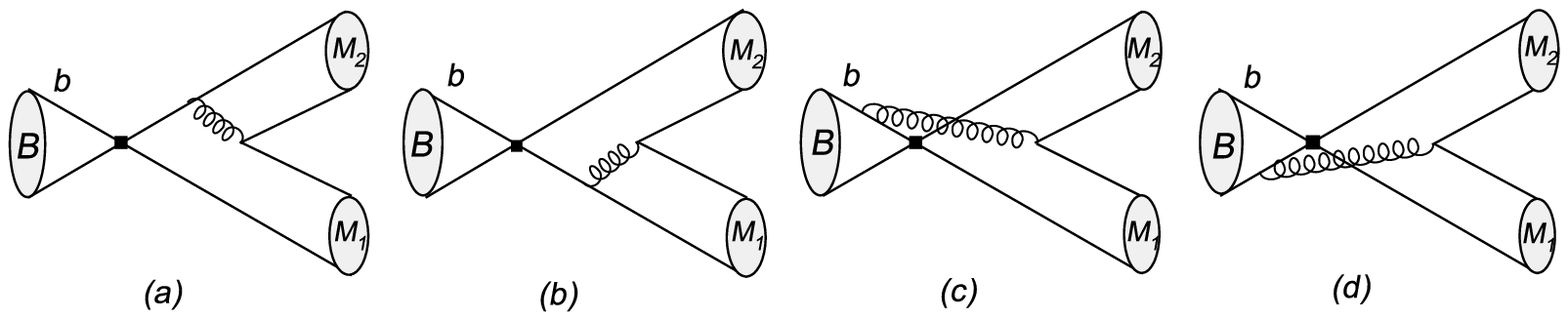}
\end{center}
\vspace{-0.6cm} \caption{\small The weak annihilation  contributions
to the coefficients $b^p_i$.}
 \label{ANN}
\end{figure}

The QCDF  \cite{BBNS} allows us to compute the nonfactorizable
corrections to the hadronic matrix elements $\langle M_1
M_2|O_i|B\rangle$ in the heavy quark limit. The  decay amplitude has
the form
\begin{eqnarray}
  \mathcal{A}^{SM}(B\to M_1M_2)
  =\frac{G_F}{\sqrt{2}}\sum_p \sum_i \lambda_p
  \Biggl\{a^p_i\langle M_2|J_2|0\rangle\langle
  M_1|J_1|B\rangle+b^p_i\langle M_1M_2|J_2|0\rangle\langle
  0|J_1|B\rangle\Biggl\},\label{AMPLITUDE}
  \end{eqnarray}
 where  the effective
parameters $a_i^p$   including nonfactorizable corrections at order
of $\alpha_s $. They are calculated from the vertex corrections, the
hard spectator scattering, and the QCD penguin contributions, which
are shown in Fig.\ref{NTL}. The parameters $b_i^p$ are calculated
from the weak annihilation contributions as shown in Fig.\ref{ANN}.

Under the naive factorization (NF) approach, the factorized matrix
element is given by
\begin{eqnarray}
A_{M_1M_2}\equiv \langle M_2|(\bar{q}_2\gamma_\mu(1-\gamma_5)q_3)
 |0\rangle
 \langle M_1|(\bar{b}\gamma^\mu(1-\gamma_5)q_1)|B \rangle.
 \end{eqnarray}
 In term of the decay constant and form
factors \cite{BallZwicky}, $A_{M_1M_2}$ are expressed as
\begin{eqnarray}
 A_{M_1M_2}=\left \{\begin{array}{ll}if_{M_2}m_B^2F_0^{B\to
 M_1}(m^2_{M_2}),
\hspace{5cm}\mbox{if $M_1=P,M_2=P$},
\\f_{M_2}m_B^2F_0^{B\to M_1}(m^2_{M_2}),
 \hspace{5.2cm}\mbox{if $M_1=P,~M_2=V$},
 \\-f_{M_2}m_B^2A_0^{B\to M_1}(m^2_{M_2}),
 \hspace{4.9cm}\mbox{if $M_1=V,~M_2=P$},
 \\-i f_{M_2}m_{M_2}\biggl[
(\varepsilon_1^{\ast}\cdot\varepsilon_2^{\ast})
(m_{B}+m_{M_1})A_1^{B\to M_1}(m_{M_2}^2) -(\varepsilon_1^{\ast}\cdot
p_B)(\varepsilon_2^{\ast}\cdot p_B)\frac{2A_2^{B \to
M_1}(m_{M_2}^2)}{m_{B}+m_{M_1}}\biggr.\\\hspace{2cm}
\left.+i\epsilon_{\mu\nu\alpha\beta}\varepsilon_2^{\ast\mu}
\varepsilon_1^{\ast\nu}p_B^{\alpha}p_1^{\beta} \frac{2V^{B\to
M_1}(m^2_{M_2})}{m_{B}+m_{M_1}}\right], \hspace{1.2cm}\mbox{if
$M_1=V,~M_2=V$},
 \end{array}
 \right.
 \end{eqnarray}
where P(V) denote a pseudoscalar(vector) meson, $p_B (m_{B})$ is the
four-momentum(mass) of the $B$ meson, $m_{M_i}$ is  the masses of
the $M_i$ mesons, and $\varepsilon_i^{\ast}$  is the polarization
vector of the vector mesons $M_i$.

 Following Beneke and Neubert
\cite{benekeNPB606}, coefficients
 $a_i^p$ can be split into two parts:
$a_i^p=a_{i,I}^p+a_{i,II}^p$. The first part contains the NF
contribution and the sum of nonfactorizable  vertex and penguin
corrections, while the second one arises from the hard spectator
scattering. The coefficients read \cite{benekeNPB606}
 \begin{eqnarray}
&& a_{1,I} =C_1+ \frac{C_2}{N_C} \left[1+\frac{C_F \alpha_s}{4\pi}
V_{M_2}\right],\qquad \qquad \qquad \qquad a_{1,II} =
\frac{C_2}{N_C} \frac{C_F \alpha_s}{4\pi}
H_{M_1M_2},\nonumber\\
&& a_{2,I} =C_2+\frac{C_1}{N_C}\left[1+\frac{C_F \alpha_s}{4\pi}
V_{M_2}\right], \qquad \qquad \qquad \qquad a_{2,II} =
\frac{C_1}{N_C} \frac{C_F \alpha_s}{4\pi}
H_{M_1M_2},\nonumber\\
&&a_{3,I}=C_3+\frac{C_4}{N_C}\left[1+\frac{C_F \alpha_s}{4\pi}
V_{M_2}\right], \qquad \qquad \qquad \qquad a_{3,II} =
\frac{C_4}{N_C} \frac{C_F \alpha_s}{4\pi}
H_{M_1M_2},\nonumber\\
&&a_{4,I}^p =C_4+\frac{C_3}{N_C}\left[1+\frac{C_F \alpha_s}{4\pi}
V_{M_2}\right]+\frac{C_F \alpha_s}{4 \pi}\frac{P^p_{M_2,2}}{N_C},
\qquad \qquad  a_{4,II} = \frac{C_3}{N_C} \frac{C_F \alpha_s}{4\pi}
H_{M_1M_2},\nonumber\\
&&a_{5,I} = C_5+\frac{C_6}{N_C}\left[1+\frac{C_F \alpha_s}{4\pi}
(-12-V_{M_2}) \right], \qquad \qquad \qquad a_{5,II} =
\frac{C_6}{N_C} \frac{C_F \alpha_s}{4\pi}
(-H_{M_1M_2}),\nonumber\\
&&a_{6,I}^p=\left\{C_6+\frac{C_5}{N_C}\left[1-6\cdot\frac{C_F
\alpha_s}{4\pi}\right]\right\}N_{M_2}+\frac{C_F \alpha_s}{4 \pi}
\frac{P^p_{M_2,3}}{N_C},~~~ a_{6,II} = 0,\nonumber\\
&& a_{7,I}=C_7+\frac{C_8}{N_C}\left[1+\frac{C_F \alpha_s}{4\pi}
(-12-V_{M_2})\right], \qquad \qquad \qquad ~ a_{7,II} =
\frac{C_8}{N_C} \frac{C_F \alpha_s}{4\pi}
(-H_{M_1M_2}),\nonumber\\
&&a_{8,I}^p =\left\{C_8+\frac{C_7}{N_C}\left[1-6\cdot\frac{C_F
\alpha_s}{4\pi}\right]\right\}N_{M_2}+\frac{\alpha_e}{9
\pi}\frac{P^{p,EW}_{M_2,3}}{N_C}, \qquad  a_{8,II} = 0,\nonumber\\
&&a_{9,I} =C_9+\frac{C_{10}}{N_C}\left[1+\frac{C_F \alpha_s}{4\pi}
V_{M_2}\right], \qquad \qquad \qquad \qquad \qquad a_{9,II} =
\frac{C_{10}}{N_C} \frac{C_F \alpha_s}{4\pi}
H_{M_1M_2},\nonumber\\
&&a_{10,I}^p=C_{10}+\frac{C_9}{N_C}\left[1+\frac{C_F \alpha_s}{4\pi}
V_{M_2}\right]+\frac{\alpha_e}{9 \pi}\frac{P^{p,EW}_{M_2,2}}{N_C},
\qquad \qquad a_{10,II} = \frac{C_9}{N_C} \frac{C_F \alpha_s}{4\pi}
H_{M_1M_2}, \label{coeffa}
 \end{eqnarray}
where $\alpha_s\equiv \alpha_s(\mu)$, $C_F=(N_C^2-1)/(2 N_C)$,
$N_C=3$ is the number of colors, and $N_{M_2}=1(0)$ for $M_2$ is a
pseudoscalar(vector) meson.  The quantities $V_{M_2}, H_{M_1M_2},
P^p_{M_2,2}, P^p_{M_2,3}$, $P^{p,EW}_{M_2,2}$ and $P^{p,EW}_{M_2,3}$
consist of convolutions of hard-scattering kernels with meson
distribution amplitudes. Specifically, the terms $V_{M_2}$ come from
the vertex corrections in Fig.\ref{NTL}(a)-\ref{NTL}(d),
$P^p_{M_2,2}$ and $P^p_{M_2,3}~ (P^{p,EW}_{M_2,2}$ and
$P^{p,EW}_{M_2,3})$ arise from QCD (EW) penguin contractions and the
contributions from the dipole operators as depicted by
Fig.\ref{NTL}(e) and \ref{NTL}(f). $H_{M_1M_2}$ is due to the hard
spectator scattering as Fig.\ref{NTL}(g) and \ref{NTL}(h). For the
penguin terms, the subscript 2 and 3 indicate the twist 2 and 3
distribution amplitudes of light mesons, respectively. Explicit
forms for these quantities are relegated to Appendix A.

We  use the convention that $M_1$ contains an antiquark from the
weak vertex, for non-singlet annihilation $M_2$ then contains a
quark from the weak vertex. The parameters $b_i^p\equiv
b^p_i(M_1,M_2)$ in Eq.(\ref{AMPLITUDE}) correspond to the weak
annihilation contributions and are given as \cite{benekeNPB675}
{\footnotesize
\begin{eqnarray}
&&b_1(M_1,M_2)=\frac{C_F}{N_C^2}C_1A^i_1(M_1,M_2), \hspace{2cm}
b_2(M_1,M_2)=\frac{C_F}{N_C^2}C_2A^i_1(M_1,M_2), \nonumber\\
&&b^p_3(M_1,M_2)=\frac{C_F}{N_C^2}\left[C_3A^i_1(M_1,M_2)
+C_5\left(A^i_3(M_1,M_2)+A^f_3(M_1,M_2)\right)
+N_CC_6A^f_3(M_1,M_2)\right],\nonumber\\
&&b^p_4(M_1,M_2)=\frac{C_F}{N_C^2}\left[C_4A^i_1(M_1,M_2)
+C_6A^i_2(M_1,M_2)\right],\nonumber\\
&&b_3^{p,ew}(M_1,M_2)=\frac{C_F}{N_C^2}\left[C_9A^i_1(M_1,M_2)
+C_7\left(A^i_3(M_1,M_2)+A^f_3(M_1,M_2)\right)
+N_CC_8A^f_3(M_1,M_2)\right],\nonumber\\
&&b_4^{p,ew}(M_1,M_2)=\frac{C_F}{N_C^2}
\left[C_{10}A^i_1(M_1,M_2)+C_8A^i_2(M_1,M_2)\right], \label{coeffb}
\end{eqnarray}}
the annihilation coefficients $(b_1, b_2), (b^p_3, b^p_4)$ and
$(b^{p,ew}_3 , b^{p,ew}_4)$ correspond to the contributions of the
tree, QCD penguins and EW penguins operators insertions,
respectively. The explicit form for the building blocks
$A^{i,f}_k(M_1,M_2)$ can be found in Appendix A.

 With the coefficients
in Eq.(\ref{coeffa}) and (\ref{coeffb}), we can obtain the decay
amplitudes of the SM part $\mathcal{A}^{SM}_f$ (the subscript ``$f$"
denotes the part without the contribution from the annihilation
part)
 and  $\mathcal{A}^{SM}_{a}$ (the subscript ``$a$"
 denotes the annihilation part). The SM part amplitudes of
 $B\to K^{(*)} \bar{K}^{(*)}$  decays are given in Appendix B.

\subsection {R-parity violating SUSY effects in the decays}

In the most general superpotential of the minimal supersymmetric
Standard Model (MSSM), the RPV superpotential is given by
\cite{RPVSW}
\begin{eqnarray}
\mathcal{W}_{\spur{R_p}}&=&\mu_i\hat{L}_i\hat{H}_u+\frac{1}{2}
\lambda_{[ij]k}\hat{L}_i\hat{L}_j\hat{E}^c_k+
\lambda'_{ijk}\hat{L}_i\hat{Q}_j\hat{D}^c_k+\frac{1}{2}
\lambda''_{i[jk]}\hat{U}^c_i\hat{D}^c_j\hat{D}^c_k, \label{rpv}
\end{eqnarray}
where $\hat{L}$ and $\hat{Q}$ are the SU(2)-doublet lepton and quark
superfields and $\hat{E}^c$, $\hat{U}^c$ and $\hat{D}^c$ are the
singlet superfields, while i, j and k are generation indices and $c$
denotes a charge conjugate field.

The bilinear RPV superpotential terms $\mu_i\hat{L}_i\hat{H}_u$ can
be rotated away by suitable redefining the lepton and Higgs
superfields \cite{barbier}. However, the rotation will generate a
soft SUSY breaking bilinear term which would affect our calculation
through penguin level. However, the processes discussed in this
paper could be induced by tree-level RPV couplings, so that we would
neglect sub-leading RPV penguin contributions in this study.

 The $\lambda$ and $\lambda'$
couplings in Eq.(\ref{rpv}) break the lepton number, while the
$\lambda''$ couplings break the baryon number. There are 27
$\lambda'_{ijk}$ couplings, 9 $\lambda_{ijk}$ and 9
$\lambda''_{ijk}$ couplings.  $\lambda_{[ij]k}$ are antisymmetric
with respect to their first two indices, and $\lambda''_{i[jk]}$ are
antisymmetric with j and k. The antisymmetry of the  baryon number
violating couplings $\lambda''_{i[jk]}$ in the last two indices
implies that there are no $\lambda''_{ijk}$ operator generating the
$\bar{b}\rightarrow \bar{s}s \bar{s}$ and $\bar{b}\rightarrow
\bar{d} d \bar{d}$ transitions.

\begin{figure}[htbp]
\begin{center}
\begin{tabular}{c}
\includegraphics[scale=0.65]{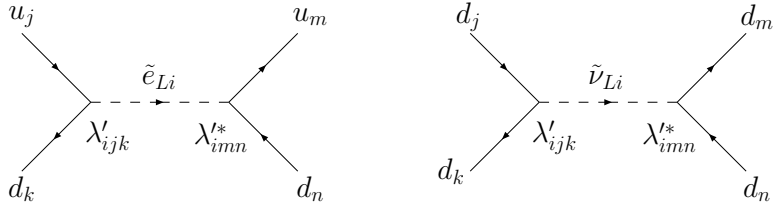}
\end{tabular}
\end{center}
\vspace{-0.6cm} \caption{\small Sleptons exchanging diagrams for
nonleptonic $B$ decays.}
 \label{Hp}
\end{figure}

\begin{figure}[htbp]
\begin{center}
\begin{tabular}{c}
\includegraphics[scale=0.65]{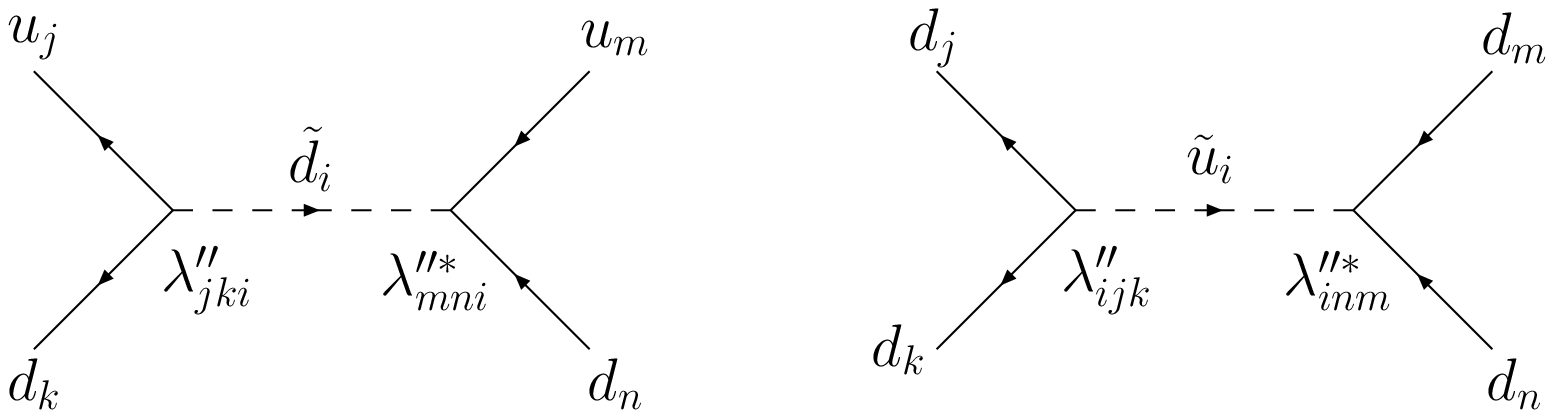}
\end{tabular}
\end{center}
\vspace{-0.6cm} \caption{\small Squarks exchanging diagrams for
nonleptonic $B$ decays.}
  \label{Hpp}
\end{figure}
From Eq.(\ref{rpv}), we can obtain the following four fermion
effective  Hamiltonian due to the sleptons exchange  as shown in
Fig.\ref{Hp}
\begin{eqnarray}
\mathcal{H}'^{\spur{R_p}}_{2u-2d}&=&\sum_i\frac{\lambda'_{ijm}
\lambda'^*_{ikl}}{2m^2_{\tilde{e}_{Li}}}
\eta^{-8/\beta_0}(\bar{d}_m\gamma^\mu P_Rd_l)_8(\bar{u}_k
\gamma_\mu P_Lu_j)_8,\nonumber\\
\mathcal{H}'^{\spur{R_p}}_{4d}&=&\sum_i
\frac{\lambda'_{ijm}\lambda'^*_{ikl}}{2m^2_{\tilde{\nu}_{Li}}}
\eta^{-8/\beta_0}(\bar{d}_m\gamma^\mu P_Rd_l)_8(\bar{d}_k\gamma_\mu
P_Ld_j)_8. \label{EqHp}
\end{eqnarray}
The four fermion effective Hamiltonian due to the squarks exchanging
as shown in Fig.\ref{Hpp} are
\begin{eqnarray}
\mathcal{H}''^{\spur{R_p}}_{2u-2d}&=&\sum_n\frac{\lambda''_{ikn}
\lambda''^*_{jln}}{2m^2_{\tilde{d}_{n}}}\eta^{-4/\beta_0}
\left\{\left[(\bar{u}_i\gamma^\mu P_Ru_j)_1(\bar{d}_k\gamma_\mu
P_Rd_l)_1-(\bar{u}_i\gamma^\mu P_Ru_j)_8(\bar{d}_k\gamma_\mu
P_Rd_l)_8\right]\right. \nonumber
\\&&\hspace{3cm}-\left[\left.(\bar{d}_k\gamma^\mu P_Ru_j)_1
(\bar{u}_i\gamma_\mu P_Rd_l)_1 -(\bar{d}_k\gamma^\mu P_Ru_j)_8
(\bar{u}_i\gamma_\mu P_Rd_l)_8\right]\right\},\nonumber\\
\mathcal{H}''^{\spur{R_p}}_{4d}&=&\sum_n
\frac{\lambda''_{nik}\lambda''^*_{njl}}{4m^2_{\tilde{u}_{n}}}
\eta^{-4/\beta_0}\left[(\bar{d}_i\gamma^\mu
P_Rd_j)_1(\bar{d}_k\gamma_\mu P_Rd_l)_1-(\bar{d}_i\gamma^\mu
P_Rd_j)_8(\bar{d}_k\gamma_\mu P_Rd_l)_8\right]. \label{EqHpp}
\end{eqnarray}
In Eqs.(\ref{EqHp}) and (\ref{EqHpp}),
$P_L=\frac{1-\gamma_5}{2},P_R=\frac{1+\gamma_5}{2},
\eta=\frac{\alpha_s(m_{\tilde{f}})}{\alpha_s(m_b)}$ and
$\beta_0=11-\frac{2}{3}n_f$. The subscript for the currents
$(j_{\mu})_{1, 8} $ represents  the current in the color singlet and
octet, respectively.  The coefficients $\eta^{-4/\beta_0}$ and
$\eta^{-8/\beta_0}$ are due to the running from the sfermion mass
scale $m_{\tilde{f}}$ (100 GeV assumed) down to the $m_b$ scale.
Since it is always assumed in phenomenology for numerical display
that only one sfermion contributes at one time, we neglect the
mixing between the operators when we use the renormalization group
equation (RGE) to run $\mathcal{H}^{\spur{R_p}}_{eff}$ down to the
low scale.

The RPV amplitude for the decays can be written as
\begin{eqnarray}
  \mathcal{A}^{\spur{R_p}}(B\to M_1M_2)&=&
  \left< M_1M_2|{\cal H}^{\spur{R_p}}_{eff}|B\right>.
\end{eqnarray}

The product RPV couplings can in general be complex and their phases
may induce new contribution to $CP$ violation, which we write as
\begin{eqnarray}
\Lambda_{ijk}\Lambda^*_{lmn} = |\Lambda_{ijk}\Lambda_{lmn}|~e^{i
\phi_{\spur{R_p}}},~~~~~\Lambda^*_{ijk}\Lambda_{lmn} =
|\Lambda_{ijk}\Lambda_{lmn}|~e^{-i \phi_{\spur{R_p}}}
\end{eqnarray}
here  the RPV coupling constant $\Lambda \in \{\lambda, \lambda',
\lambda''\}$, and $\phi_{\spur{R_p}}$ is the RPV weak phase, which
may be any value between $-\pi$ and $\pi$.

For simplicity we only consider the vertex corrections and the hard
spectator scattering in the RPV decay amplitudes.  We ignore the RPV
penguin contributions,
  which are expected to be small even compared with the SM penguin
   amplitudes,  this follows from the smallness of the relevant RPV
couplings compared with the SM gauge couplings. The bounds on the
RPV couplings are insensitive to the inclusion of the RPV penguins
\cite{RPVpipi}. We also neglected the annihilation contributions in
the RPV amplitudes. The R-parity violating part of  the decay
amplitudes $\mathcal{A}^{\spur{R_p}}$  can be found  in Appendix C.

\subsection{The total decay amplitude}
With the QCDF, we can get the total decay amplitude
\begin{eqnarray}
\mathcal{A}(B\rightarrow M_1 M_2)=\mathcal{A}^{SM}_f(B\rightarrow
M_1 M_2)+\mathcal{A}^{SM}_a(B\rightarrow M_1
M_2)+\mathcal{A}^{\spur{R_p}}(B\rightarrow M_1 M_2). \label{amp}
\end{eqnarray}
The expressions for the SM amplitude $\mathcal{A}^{SM}_{f,a} $ and
the RPV amplitude $ \mathcal{A}^{\spur{R_p}}$ are presented in
Appendices B and C, respectively. From the amplitude in
Eq.(\ref{amp}), the branching ratio reads
\begin{eqnarray}
\mathcal{B}(B\rightarrow M_1 M_2)=\frac{\tau_B |p_c |}{8\pi
m_B^2}\left|\mathcal{A}(B\rightarrow M_1 M_2)\right|^2S,
\end{eqnarray}
where $S=1/2$  if  $M_1$ and $M_2$ are identical, and $S = 1$
otherwise; $\tau_B$ is the B lifetime, $|p_c|$ is the center of mass
momentum  in the center of mass frame of $B$ meson, and given by
\begin{eqnarray}
|p_c|=\frac{1}{2m_B}\sqrt{[m_B^2-(m_{M_1}+m_{M_2})^2][m_B^2
-(m_{M_1}-m_{M_2})^2]}.
\end{eqnarray}

 The direct  $CP$ asymmetry is defined as
\begin{eqnarray}
\mathcal{A}_{CP}^{dir}=\frac{\mathcal{B}(\bar{B}\to
\bar{f})-\mathcal{B}(B\to f)}{\mathcal{B}(\bar{B}\to
\bar{f})+\mathcal{B}(B\to f)}.
\end{eqnarray}

In the $B\to VV$ decay, the longitudinal polarization fraction is
defined by
\begin{eqnarray}
f_{L}=\frac{\Gamma_L}{\Gamma}=\frac{|\mathcal{A}_0|^2}
{|\mathcal{A}_0|^2+|\mathcal{A}_+|^2+|\mathcal{A}_-|^2},
\end{eqnarray}
where $A_0(A_{\pm})$ corresponding to the longitudinal(two
transverse) polarization  amplitude(s) for $B\to VV$ decay.

\section{Input Parameters}
{\bf A. Wilson coefficients} \\
We use the next-to-leading  Wilson coefficients calculated  in the
naive dimensional regularization (NDR) scheme  at $m_b$ scale
 \cite{coeff}:
\begin{eqnarray}
&&C_1=1.082,~~~C_2=-0.185,~~~C_3=0.014,~~~C_4=-0.035,
~~~C_5=0.009,\nonumber\\
&&C_6=-0.041,~~~ C_7/\alpha_e =-0.002,~~~C_8/\alpha_e
=0.054,~~~C_9/\alpha_e
=-1.292, \nonumber \\
&&C_{10}/\alpha_e =0.263,~~~
C^{eff}_{7\gamma}=-0.299,~~~C^{eff}_{8g}=-0.143.
\end{eqnarray}
{\bf B. The CKM matrix element} \\
The magnitude of the CKM elements are taken from \cite{PDG}:
 \begin{equation}
 \begin{array}{lll}
 |V_{ud}|=0.9738\pm0.0005, & |V_{us}|=0.2200\pm0.0026,
 & |V_{ub}|=0.00367\pm0.00047,\\
 |V_{cd}|=-0.224\pm0.012, & |V_{cs}|=0.996\pm0.013,
 & |V_{cb}|=0.0413\pm0.0015,\\
| V_{tb}^*V_{td}|=0.0083\pm0.0016 & |V_{tb}V_{ts}^*|=-0.047\pm0.008, & \\
  \end{array}
 \end{equation}
and the CKM phase $\gamma=60^\circ \pm 14^\circ$,
sin$(2\beta)=0.736\pm 0.049$.\\
{\bf C. Masses and lifetime }\\
 There are two types of quark mass in our analysis. One type is
 the pole mass which appears
in the loop integration. Here we fix them as
\begin{eqnarray}
m_u=m_d=m_s=0,~~~~m_c=1.47 ~GeV,~~~~m_b=4.8 ~GeV.
\end{eqnarray}
The other type quark mass appears in the hadronic matrix elements
and the chirally enhanced factor
$r_\chi^P=\frac{2\mu_p}{\overline{m}_b}$ through the equations of
motion. They are renormalization scale dependent. We shall use  the
2004 Particle Data Group data \cite{PDG} for discussion:
\begin{eqnarray}
&&\overline{m}_u(2GeV)=0.0015\sim 0.004~
GeV,~~~~\overline{m}_d(2GeV)=0.004\sim0.008~
GeV,\nonumber\\
 &&\overline{m}_s(2GeV)=0.08\sim 0.13 ~
GeV,~~~~~~~~~\overline{m}_b(\overline{m}_b)=4.1\sim4.4~ GeV,
\end{eqnarray}
and then employ the formulae in Ref. \cite{coeff}
\begin{eqnarray}
\overline{m}(\mu)=\overline{m}(\mu_0)\left[\frac{\alpha_s(\mu)}
{\alpha_s(\mu_0)}\right]^{\frac{\gamma^{(0)}_m}{2
\beta_0}}
\left[1+\left(\frac{\gamma^{(1)}_m}{2\beta_0}
-\frac{\beta_1\gamma^{(0)}_m}{2\beta_0^2}\right)
\frac{\alpha_s(\mu)-\alpha_s(\mu_0)}{4 \pi}\right],
\end{eqnarray}
to obtain the current quark masses to any scale. The definitions of
$\gamma^{(0)}_m, \gamma^{(1)}_m, \beta_0, \beta_1$ can be found in
\cite{coeff}.

To compute the branching ratio, the  masses of meson are also taken
from \cite{PDG}
\begin{center}
\begin{tabular}{cccc}
$m_{B_{u}}=5.279~GeV$,&$m_{K^{*\pm}}=0.892~GeV$,&$m_{K^{\pm}}=0.494~GeV$,\\
$m_{B_{d}}=5.279~GeV$,&$m_{K^{*0}}=0.896~GeV$,&$m_{K^{0}}=0.498~GeV$.
\end{tabular}
\end{center}

The lifetime of $B$ meson \cite{PDG}
\begin{eqnarray}
\tau_{B_{u}}=(1.638\pm 0.011)~ps,~~\tau_{B_{d}}=(1.532\pm 0.009)~ps.
\end{eqnarray}
{\bf D. The LCDAs of the  meson}\\
 For the LCDAs of the
 meson, we use the asymptotic form \cite{M
braun,DAchernyak,pallbraun}
\begin{eqnarray}
\Phi_P(x)=6x(1-x),~~~~\Phi^P_p(x)=1,
\end{eqnarray}
 for the pseudoscalar meson, and
\begin{eqnarray}
&&\Phi_{\parallel}^V (x)=\Phi_{\perp}^V
(x)=g_{\perp}^{(a)V}=6x(1-x),\nonumber\\
&& g_{\perp}^{(v)V} (x)=\frac{3}{4}[1+(2x-1)^2],
\end{eqnarray}
 for the vector meson.

 We adopt
the moments of the $\Phi_1^B (\xi)$ defined in Ref.
\cite{BBNS,benekeNPB675} for our numerical evaluation:
\begin{equation}
\int_0^1{\rm d}\xi\frac{\Phi_1^B
(\xi)}{\xi}=\frac{m_{B}}{\lambda_B},
\end{equation}
with $\lambda_B=(0.46\pm0.11)~GeV$ \cite{lamdB}. The quantity
$\lambda_B$ parameterizes our ignorance about the $B$ meson
distribution amplitudes and thus brings considerable theoretical
uncertainty.\\
 {\bf E. The decay constants and form factors} \\For
the decay constants, we take the latest light-cone QCD sum rule
results (LCSR) \cite{BallZwicky} in our calculations:
\begin{eqnarray}
f_{B_{u(d)}}=0.161~GeV,~~f_{K}=0.160~GeV,~~f_{K^*}=0.217~GeV,
~~f^{\perp}_{K^*}=0.156~GeV.
\end{eqnarray}
For the form factors involving the $B\to K^{(*)}$
transition, we adopt the the values given by \cite{BallZwicky}
 \begin{eqnarray}
  &&A^{B_{u(d)}\to
  K^*}_{0}(0)=0.374\pm0.034,~~~
    A_1^{B_{u(d)}\rightarrow K^{\ast}}(0)=0.292\pm 0.028,\nonumber\\
 &&A_2^{B_{u(d)}\rightarrow K^{\ast}}(0)=0.259\pm 0.027,~~~
 V^{B_{u(d)}\rightarrow K^{\ast}} (0) =0.411\pm 0.033,\\
&& F^{B_{u(d)}\to K}_{0}(0)=0.331\pm 0.041.\nonumber
 \end{eqnarray}

\section{Numerical results and Analysis}
First, we will show our estimations in the SM by taking the center
value of the input parameters and compare with the relevant
experimental data. Then, we will consider the RPV effects to
constrain the relevant RPV couplings from  the experimental data.
Using the constrained parameter spaces, we will give the RPV SUSY
predictions for the branching ratios, the direct $CP$ asymmetries
and the longitudinal polarizations, which have not been measure yet
in $B\to K^{(*)}\bar{K}^{(*)}$ system.

When considering  the RPV effects, we will use the input parameters
and the experimental data which are varied randomly within $1\sigma$
variance.  In the SM, the weak phase $\gamma$ is well constrained,
however, with the presence of the RPV, this constraint may be
relaxed. We would not take $\gamma$ within the SM range, but vary it
randomly in the range of 0 to $\pi$ to obtain conservative limits on
RPV couplings. We assume that only one sfermion contributes at one
time with a mass of 100 GeV.  As for other values of the sfermion
masses, the bounds on the couplings in this paper can be easily
obtained by scaling them with factor $\tilde{f}^2\equiv
(\frac{m_{\tilde{f}}}{100GeV})^2$.

For the $B \rightarrow K^{(*)} \bar{K}^{(*)}$ modes, several
branching ratios and one direct $CP$ asymmetry have been measured by
\textit{BABAR}, Belle and CLEO \cite{PDG,expdata}, and their
averaged values \cite{HFAG} are
\begin{eqnarray}
\mathcal{B}(B^+_u \rightarrow K^+ \bar{K}^{0})&
=&(1.2\pm0.3)\times10^{-6}, \nonumber\\
\mathcal{B}(B^0_d \rightarrow K^0 \bar{K}^{0})&
=&(0.96^{+0.25}_{-0.24})\times10^{-6}, \nonumber\\
\mathcal{B}(B^+_u \rightarrow K^+ \bar{K}^{*0})
&<&5.3\times10^{-6}~~~~(~90\%~CL~),\nonumber\\
\mathcal{B}(B^+_u \rightarrow K^{*+} \bar{K}^{*0})
&<&71\times10^{-6}~~~~(~90\%~CL~), \nonumber\\
\mathcal{B}(B^0_d \rightarrow K^{*0}\bar{K}^{*0})
&<&22\times10^{-6}~~~~(~90\%~CL~), \nonumber\\
\mathcal{A}^{dir}_{CP}(B^+_u \rightarrow
K^{+}\bar{K}^0)&=&0.15\pm0.33. \label{data}
\end{eqnarray}

The numerical results in the SM  are presented in Table I, which
shows the results for the $CP$ averaged branching ratios
($\mathcal{B}$), the direct $CP$ asymmetries
($\mathcal{A}^{dir}_{CP}$) and the longitudinal polarization
fractions ($f_L$).
\begin{table}[htbp]
\centerline{\parbox{11.6cm}{{Table I: The SM predictions   for
$\mathcal{B}$ (in unit of $10^{-6}$), $\mathcal{A}^{dir}_{CP}$ and
$f_L$ in $B\to K^{(*)}\bar{K}^{(*)}$ decays in the framework of NF
and QCDF.}}} \vspace{0.4cm}
\begin{center}
\begin{tabular}
{l|cc|cr|cc}\hline\hline
 \multicolumn{2}{c@{\hspace{-9.03cm}}}{\vline$\hspace{1.24cm}
 \mathcal{B}\hspace{1.24cm}
 \vline \hspace{0.98cm}\mathcal{A}^{dir}_{CP}\hspace{0.98cm}
 \vline\hspace{1.2cm}f_L$}\\
\cline{2-7}\raisebox{2.3ex}[0pt]{
 ~~~Decays} & NF&QCDF& NF&QCDF&NF&QCDF\\\hline
 $B^+_u \rightarrow K^+ \bar{K}^{0}$&0.61&0.89&0.00&-0.13&&\\
$B^0_d \rightarrow K^0 \bar{K}^{0}$&0.57&0.89&0.00&-0.13&&\\
$B^+_u \rightarrow K^{*+}\bar{K}^0$&0.06&0.10&0.00&-0.19&&\\
$B^+_u \rightarrow K^+ \bar{K}^{*0}$&0.15&0.18&0.00&-0.08&&\\
$B^0_d \rightarrow K^{*0}\bar{K}^0$&0.05&0.10&0.00&-0.18&&\\
$B^0_d \rightarrow K^0 \bar{K}^{*0}$&0.14&0.16&0.00&-0.10&&\\
$B^+_u \rightarrow K^{*+} \bar{K}^{*0}$&0.20&0.22&0.00&-0.22&0.91&0.90\\
$B^0_d \rightarrow K^{*0}\bar{K}^{*0}$&0.19&0.20&0.00&-0.22&0.91&0.90\\
\hline\hline
\end{tabular}
\end{center}
\end{table}

From Table I, we can see that the branching ratios for them are
expected to be quite small, of order $10^{-7}$, since $B\to
K^{(*)}\bar{K}^{(*)}$ are the pure $b\to d$ penguin dominated
decays. The subleading diagrams may lead to the significant $CP$
violations in the most  $B\to K^{(*)}\bar{K}^{(*)}$ decays. As
$B^0_d\to K^{\pm} K^{\mp}$ decays involved only non-factorizable
annihilation contributions, their branching ratios are much smaller
than those of $B \rightarrow K^+ \bar{K}^{0},K^0\bar{K}^0$ decays,
we would not study the $B^0_d\to K^{\pm}K^{\mp}$ modes in this
paper. It should be noted  that the amplitude for $\bar{B}^0_d\to
K^0 \bar{K}^{*0}$ is not simply related to that for $B^0_d\to K^0
\bar{K}^{*0}$ since the spectator quark is part of the $K^0$  in the
latter decay, while in the former in the $\bar{K}^{*0}$.

\begin{figure}[h]
\begin{center}
\includegraphics[scale=2.2]{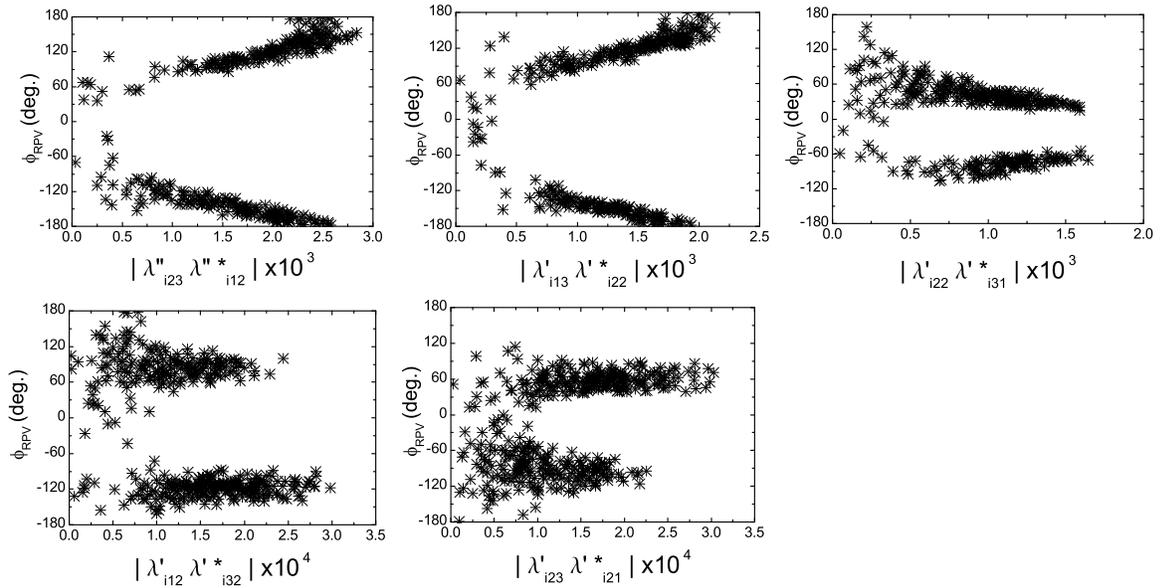}
\end{center}
\vspace{-0.6cm}
 \caption{ The allowed parameter spaces for the relevant
 RPV couplings constrained by $B\to K^{(*)} \bar{K}^{(*)}$, and $\phi_{RPV}$ denotes the RPV weak phase.}
 \label{bounds}
\end{figure}

Although recent experimental results in $B\to K^{(*)} \bar{K}^{(*)}$
seem to be roughly consistent with the SM predictions, there are
still windows for  NP in these processes.
 We now turn to the  RPV effects  in $B\to K^{(*)} \bar{K}^{(*)}$ decays.
  There are five RPV coupling constants
contributing to the eight  $B\to K^{(*)} \bar{K}^{(*)}$ decay modes.
We use $\mathcal{B}$, $\mathcal{A}^{dir}_{CP}$ and the
 experimental constraints shown in Eq.(\ref{data}) to
constrain the relevant RPV parameters. As known, data on low energy
processes can be used to impose rather strictly constraints on many
of these couplings. In  Fig.\ref{bounds}, we present the bounds on
the RPV couplings.  The random variation of the parameters
subjecting to the constraints as discussed above leads to the
scatter plots displayed in  Fig.\ref{bounds}.

From Fig.\ref{bounds}, we find that every RPV weak phase has two
possible bands, one band is for positive value of RPV weak phase,
and another for negative one.  We also find  the magnitudes of the
relevant RPV couplings have been up limited. The upper limits are
summarized in Table II. For comparison,  the existing bounds on
these quadric coupling products \cite{report,hexg} are also listed.
Our bounds on $|\lambda'_{i13}\lambda'^*_{i22}|$,
$|\lambda'_{i12}\lambda'^*_{i32}|$ and
$|\lambda'_{i23}\lambda'^*_{i21}|$ are stronger than the existing
ones.
\begin{table}[htbp]
\centerline{\parbox{14.1cm}{Table II: Bounds for  the relevant RPV
couplings  by $B \to K^{(*)} \bar{K}^{(*)}$ decays for 100 GeV
sfermions and previous bounds  are  listed for comparison. }}
\vspace{0.5cm}
\begin{center}
\begin{tabular}{c|l|l}\hline\hline
Couplings&~~~~~~~~~Bounds [Process]& Previous bounds [Process]\\
\hline $|\lambda''_{i23}\lambda''^*_{i12}|$&$\leq2.9\times
10^{-3}~[B\to K^{(*)}\bar{K}^{(*)}$] &$^{\leq 5.\times 10^{-3}~[B\to
K \bar{K}]}_{\leq 6.\times 10^{-5}
~[B^0\to \phi \pi^0,\phi\phi]}$ \cite{report}\\
$|\lambda'_{i13}\lambda'^*_{i22}| $&$\leq2.2\times 10^{-3}~[
B\to K^{(*)}\bar{K}^{(*)}]$&$\leq2.9\times 10^{-3}[B\to K\bar{K}]$
 \cite{hexg}\\
$|\lambda'_{i22}\lambda'^*_{i31}|$&$\leq1.7\times 10^{-3}~[B\to
K^{(*)}\bar{K}^{(*)}]$&$\leq1.\times 10^{-4}~[K \bar{K}]$\cite{report}\\
$|\lambda'_{i12}\lambda'^*_{i32}| $&$\leq3.0\times 10^{-4}~[B\to
K\bar{K}^{(*)},\bar{K}K^{(*)}]$&$\leq4.\times 10^{-4}~[B^0\to \phi
\pi^0]$ \cite{report}\\
$|\lambda'_{i23}\lambda'^*_{i21}| $&$\leq3.0\times 10^{-4}~[B\to
K\bar{K}^{(*)},\bar{K}K^{(*)}]$&$\leq4.\times 10^{-4}~[B^0\to \phi
\pi^0]$ \cite{report}\\\hline\hline
\end{tabular}
\end{center}
\end{table}
\begin{table}[htbp]\centerline{\parbox{16.5cm}{{Table III: The
theoretical predictions for $\mathcal{B}$ (in unit of $10^{-6}$),
$\mathcal{A}^{dir}_{CP}$ and $f_L$ base on the RPV SUSY model, which
are obtained by the allowed regions of the different RPV couplings.
}}} \vspace{0.6cm}
\begin{center}\small{
\begin{tabular}
{l|l|l|l|l|l}\hline\hline
&$\lambda''_{i23}\lambda''^*_{i12}$&$\lambda'_{i13}\lambda'^*_{i22}$
&$\lambda'_{i22}\lambda'^*_{i31}$
&$\lambda'_{i12}\lambda'^*_{i32}$
&$\lambda'_{i23}\lambda'^*_{i21}$\\\hline $\mathcal{B}(B^+_u
\rightarrow K^{*+}\bar{K}^0)$&$[0.0052,7.8]$
&$[0.013,5.5]$&$[0.0059,6.4]$ &$[0.056,1.4]$&$[0.064,1.3]$\\
$\mathcal{B}(B^+_u \rightarrow K^+\bar{K}^{*0})$
&$[0.071,5.3]$&$[0.056,5.3]$&$[0.0096,5.3]$ &&\\
$\mathcal{B}(B^0_d \rightarrow K^{*0}\bar{K}^0)$
&$[0.0060,7.5]$&$[0.011,5.1]$&$[0.0053,6.1]$
&$[0.049,1.5]$&$[0.054,1.2]$\\
$\mathcal{B}(B^0_d \rightarrow K^0\bar{K}^{*0})$
&$[0.069,5.0]$&$[0.050,5.1]$&$[0.0093,5.0]$ &&\\
$\mathcal{B}(B^+_u \rightarrow K^{*+}\bar{K}^{*0})$
&$[0.087,19]$&$[0.041,23]$&$[0.029,16]$ &&\\
$\mathcal{B}(B^0_d \rightarrow K^{*0}\bar{K}^{*0})$ &$[0.080,17]$
&$[0.039,22]$&$[0.027,15]$ &&\\\hline
 $\mathcal{A}^{dir}_{CP}(B^0_d
\rightarrow K^0\bar{K}^{0})$
&$[-0.75,0.57]$&$[-0.19,0.44]$&$[-0.18,0.47]$
&$[-0.18,0.47]$&$[-0.18,0.50]$\\
$\mathcal{A}^{dir}_{CP}(B^+_u \rightarrow K^{*+}\bar{K}^{0})$
&$[-0.19,0.17]$&$[-0.32,0.17]$&$[-0.42,0.47]$
&$[-0.99,0.99]$&$[-0.98,0.76]$\\
$\mathcal{A}^{dir}_{CP}(B^+_u \rightarrow K^+
\bar{K}^{*0})$&$[-0.63,0.63]$&$[-0.38,0.47]$&$[-0.65,0.38]$ &&\\
$\mathcal{A}^{dir}_{CP}(B^0_d \rightarrow
K^{*0}\bar{K}^0)$&$[-0.28,0.19]$
&$[-0.33,0.17]$&$[-0.28,0.80]$ &$[-0.99,0.99]$&$[-0.99,0.73]$\\
$\mathcal{A}^{dir}_{CP}(B^0_d \rightarrow K^0
\bar{K}^{*0})$&$[-0.76,0.62]$&$[-0.38,0.48]$&$[-0.39,0.40]$ &\\
$\mathcal{A}^{dir}_{CP}(B^+_u \rightarrow K^{*+}
\bar{K}^{*0})$&$[-0.63,0.30]$&$[-0.26,0.25]$&$[-0.77,0.32]$ &&\\
$\mathcal{A}^{dir}_{CP}(B^0_d \rightarrow
K^{*0}\bar{K}^{*0})$&$[-0.46,0.38]$ &$[-0.26,0.25]$&$[-0.77,0.32]$
&&\\\hline $f_L(B^+_u \rightarrow K^{*+}
\bar{K}^{*0})$&$[0.72,0.97]$ &$[0.59,0.95]$&$[0.74,0.93]$&&\\
$f_L(B^0_d \rightarrow K^{*0}\bar{K}^{*0})$&$[0.72,0.97]$
&$[0.59,0.95]$&$[0.74,0.93]$ &&\\ \hline\hline
\end{tabular}}
\end{center}
\end{table}

Using the constrained  parameter spaces shown in Fig.\ref{bounds},
one can predict the RPV effects on the other quantities  which have
not been measured yet in $B\to K^{(*)} \bar{K}^{(*)}$ decays. With
the expressions for $\mathcal{B}$, $\mathcal{A}^{dir}_{CP}$ and
$f_L$ at hand, we perform a scan on the input parameters and the new
constrained RPV coupling spaces. Then the allowed ranges for
$\mathcal{B}$, $\mathcal{A}^{dir}_{CP}$ and $f_L$ are obtained with
five different RPV couplings, which satisfy all present experimental
constraints shown in Eq.(\ref{data}).

We obtain that the RPV effects could  alter the predicted
$\mathcal{B}$ and $\mathcal{A}^{dir}_{CP}$  significantly from their
SM values. For decay modes, which have not been measured yet, their
branching ratios  can be changed one or two order(s) of magnitude
comparing with the SM expectations,
\begin{eqnarray}
&&9.\times 10^{-9}<\mathcal{B}(B\rightarrow
K^+\bar{K}^{*0},K^0\bar{K}^{*0})< 5.\times 10^{-6}, \nonumber\\
&&5.\times 10^{-9}<\mathcal{B}(B\rightarrow
K^{*+}\bar{K}^{0},K^{*0}\bar{K}^{0})< 8.\times 10^{-6}, \nonumber\\
&&3.\times 10^{-8}<\mathcal{B}(B\rightarrow
K^{*+}\bar{K}^{*0},K^{*0}\bar{K}^{*0})< 2.\times 10^{-5},
\end{eqnarray}
especially, the upper limit of $\mathcal{B}(B\rightarrow
K^{*+}\bar{K}^{*0})$ $<2.\times 10^{-5}$ which we have obtained is
smaller than the experimental upper limit  $<7.\times 10^{-5}$. For
$\mathcal{A}^{dir}_{CP}$, the RPV predictions  on two decays
$B\rightarrow K^{*+}\bar{K}^{*0},K^{*0}\bar{K}^{*0}$ are
\begin{eqnarray}
\mathcal{A}^{dir}_{CP}(B\rightarrow K^{*+}\bar{K}^{*0})\leq
0.32,~~~~~~~\mathcal{A}^{dir}_{CP}(B\rightarrow
K^{*0}\bar{K}^{*0})\leq 0.38,
\end{eqnarray}
and there are quite loose constraints on the direct CP asymmetries
of  the other five decays $B\rightarrow
K^{0}\bar{K}^{0},K^{*+}\bar{K}^{0},K^{+}\bar{K}^{*0},K^{*0}\bar{K}^{0},K^{0}\bar{K}^{*0}$.
But the RPV effects on the $f_L(B\rightarrow
K^{*+}\bar{K}^{*0},K^{*0}\bar{K}^{*0})$ are found to be very small,
$f_L(B\rightarrow K^{*+}\bar{K}^{*0},K^{*0}\bar{K}^{*0})$ are found
to lie between 0.7 and 1, and these intervals  are mainly due to the
parameter uncertainties not  the RPV effects. So we might come to
the conclusion, the RPV SUSY predictions show that the decays
$B\rightarrow K^{*+}\bar{K}^{*0},K^{*0}\bar{K}^{*0}$ are dominated
by the longitudinal polarization,  and there are not abnormal large
transverse  polarizations in $B_{u,d}\to K^{*} \bar{K}^{*}$ decays.
The detailed numerical ranges which obtained by different RPV
couplings are summarized in Table III.

In Figs.\ref{li23ppli12pp}-\ref{li23pli21p}, we present correlations
between the physical observable $\mathcal{B}$,
$\mathcal{A}^{dir}_{CP}$, $f_L$ and the parameter spaces of
different RPV couplings by the three-dimensional scatter plots.
 The more information are displayed in
 Figs.\ref{li23ppli12pp}-\ref{li23pli21p}, we can see the change trends of
the physical observable quantities with  the modulus and weak phase
$\phi_{\spur{R_p}}$ of RPV couplings. We take the first plot in
Fig.\ref{li23ppli12pp} as an example, this plot shows that
$\mathcal{B}(B\rightarrow K^{*+}\bar{K}^{0})$ change trend with  RPV
coupling $\lambda''_{i23}\lambda''^*_{i12}$. We also give
projections on three vertical
 planes, the $|\lambda''_{i23}
 \lambda''^*_{i12}|$-$\phi_{\spur{R_p}}$ plane display the allowed
 regions of $\lambda''_{i23}
 \lambda''^*_{i12}$ which satisfy experimental data in
 Eq.(\ref{data}) (the same as the first plot in Fig.\ref{bounds}).
  It's shown that $\mathcal{B}(B\rightarrow
 K^{*+}\bar{K}^{0})$ is increasing  with $|\lambda''_{i23}
 \lambda''^*_{i12}|$ on the $\mathcal{B}(B\rightarrow K^{*+}\bar{K}^{0})$
 -$|\lambda''_{i23} \lambda''^*_{i12}|$ plane.  From the $\mathcal{B}(B\rightarrow
 K^{*+}\bar{K}^{0})$-$\phi_{\spur{R_p}}$ plane,   we can see that $\mathcal{B}(B\rightarrow
 K^{*+}\bar{K}^{0})$ is increasing with  $|\phi_{\spur{R_p}}|$.
  Further refined measurements of $\mathcal{B}(B\rightarrow
  K^{*+}\bar{K}^{0})$ can further restrict the constrained space of $\lambda''_{i23}
 \lambda''^*_{i12}$, whereas with more narrow space of $\lambda''_{i23}
 \lambda''^*_{i12}$  more accurate $\mathcal{B}(B\rightarrow
  K^{*+}\bar{K}^{0})$ can be predicted.

  The following salient features
  in Figs.\ref{li23ppli12pp}-\ref{li23pli21p} are summarized as following.

\begin{itemize}

\item  Fig.\ref{li23ppli12pp} displays the effects of RPV coupling $\lambda''_{i23}
 \lambda''^*_{i12}$ on $\mathcal{B}$,
$\mathcal{A}^{dir}_{CP}$ and $f_L$ in $B\to K^{(*)}\bar{K}^{(*)}$.
The constrained $|\lambda''_{i23}
 \lambda''^*_{i12}|$-$\phi_{\spur{R_p}}$ plane shows the
 allowed range of $\lambda''_{i23}
 \lambda''^*_{i12}$ as in the first plot of
 Fig.\ref{bounds}. The six
  $\mathcal{B}(B\to K^{*+}\bar{K}^{0},K^{+}\bar{K}^{*0},
  K^{*0}\bar{K}^{0},K^{0}\bar{K}^{*0},K^{*+}\bar{K}^{*0},K^{*0}\bar{K}^{*0})$
  have the similar change trends with $|\lambda''_{i23}
 \lambda''^*_{i12}|$ and $|\phi_{\spur{R_p}}|$, and they are increasing  with
 $|\lambda''_{i23} \lambda''^*_{i12}|$ and $|\phi_{\spur{R_p}}|$.
 $|\mathcal{A}^{dir}_{CP}(B\to K^{0}\bar{K}^{0})|$
 are increasing with $|\phi_{\spur{R_p}}|$, but $|\lambda''_{i23}
 \lambda''^*_{i12}|$ has small effect on
 $\mathcal{A}^{dir}_{CP}(B\to K^{0}\bar{K}^{0})$.
The two
 $|\mathcal{A}^{dir}_{CP}(B\to K^{+}\bar{K}^{*0},
 K^{0}\bar{K}^{*0})|$ tend
 to zero with increasing  $|\lambda''_{i23}
 \lambda''^*_{i12}|$ and $|\phi_{\spur{R_p}}|$.
 The other four
 $|\mathcal{A}^{dir}_{CP}(B\to K^{*+}\bar{K}^{0},K^{*0}\bar{K}^{0},
 K^{*+}\bar{K}^{*0},K^{*0}\bar{K}^{*0})|$ tend
 to zero with increasing $|\phi_{\spur{R_p}}|$,
 and they could have smaller ranges with $|\lambda''_{i23}
 \lambda''^*_{i12}|$.
 The RPV effects on the $f_L(B\rightarrow
K^{*+}\bar{K}^{*0},K^{*0}\bar{K}^{*0})$ are very small, and
$f_L(B\rightarrow K^{*+}\bar{K}^{*0},K^{*0}\bar{K}^{*0})$ are found
to lie between 0.72 and 0.97.

\item The effects of $\lambda'_{i13}
 \lambda'^*_{i22}$ on $\mathcal{B}$,
$\mathcal{A}^{dir}_{CP}$ and $f_L$ are exhibited in
Fig.\ref{li13pli22p}. The constrained $|\lambda'_{i13}
 \lambda'^*_{i22}|$-$\phi_{\spur{R_p}}$ plane is the same as the second plot
 in  Fig.\ref{bounds}. The effects of $\lambda'_{i13}
 \lambda'^*_{i22}$ on $\mathcal{B}$,
$\mathcal{A}^{dir}_{CP}$ and $f_L$ are similar to $\lambda''_{i23}
 \lambda''^*_{i12}$ shown in Fig.\ref{li23ppli12pp}.

\item In Fig.\ref{li22pli31p}, we plot $\mathcal{B}$,
$\mathcal{A}^{dir}_{CP}$ and $f_L$ as functions of $\lambda'_{i22}
 \lambda'^*_{i31}$. The constrained $|\lambda'_{i22}
 \lambda'^*_{i31}|$-$\phi_{\spur{R_p}}$ plane is the same  as the third plot of
 Fig.\ref{bounds}. The six
  branching ratios are increasing  with
 $|\lambda'_{i22} \lambda'^*_{i31}|$ and  decreasing with $|\phi_{\spur{R_p}}|$.
 $|\mathcal{A}^{dir}_{CP}(B\to K^{0}\bar{K}^{0})|$
 is unaffected  by  $|\lambda'_{i22}
 \lambda'^*_{i31}|$, but the other six direct CP asymmetries
 could have smaller ranges with $|\lambda'_{i22}
 \lambda'^*_{i31}|$.
 $|\mathcal{A}^{dir}_{CP}(K^{*+}\bar{K}^{0},K^{*0}\bar{K}^{0})|$
tends to zero with decreasing  $|\phi_{\spur{R_p}}|$, however,
 $\phi_{\spur{R_p}}$ has small effect on
 $\mathcal{A}^{dir}_{CP}(B\to K^{0}\bar{K}^{0},K^{+}\bar{K}^{*0},
K^{0}\bar{K}^{*0}, K^{*+}\bar{K}^{*0},K^{*0}\bar{K}^{*0})$. The
$\lambda'_{i22}
 \lambda'^*_{i31}$ effects on
the $f_L(B\rightarrow K^{*+}\bar{K}^{*0},K^{*0}\bar{K}^{*0})$ are
 small.

\item  RPV coupling $\lambda'_{i12}
 \lambda'^*_{i32}$  contributes to the  decays
 $B\to K^{+}\bar{K}^{0},K^{0}\bar{K}^{0},
 K^{*+}\bar{K}^{0},K^{*0}\bar{K}^{0}$, and the effects
 are shown in Fig.\ref{li12pli32p}. The constrained $|\lambda'_{i12}
 \lambda'^*_{i32}|$-$\phi_{\spur{R_p}}$ plane is the same
 as the fourth  plot in Fig.\ref{bounds}. We can see that
  $\mathcal{B}(B\to K^{*+}\bar{K}^{0},K^{*0}\bar{K}^{0})$
  are rising  with $|\lambda'_{i12} \lambda'^*_{i32}|$,
  and unaffected by $\phi_{\spur{R_p}}$.
   $\mathcal{A}^{dir}_{CP}(B\to K^{0}\bar{K}^{0})$
  is steady  against  $|\lambda'_{i12} \lambda'^*_{i32}|$, and
  $|\mathcal{A}^{dir}_{CP}(B\to K^{*+}\bar{K}^{0},K^{*0}\bar{K}^{0})|$
 could have smaller ranges with $|\lambda'_{i12} \lambda'^*_{i32}|$.
  $\mathcal{A}^{dir}_{CP}(B\to K^{0}\bar{K}^{0},K^{*+}
   \bar{K}^{0},$ $K^{*0}\bar{K}^{0})$
  are becoming large  with increasing  of   $|\phi_{\spur{R_p}}|$.

\item  $\lambda'_{i23}
 \lambda'^*_{i21}$  also only contributes to the  decays
 $B\to K^{+}\bar{K}^{0},K^{0}\bar{K}^{0},
 K^{*+}\bar{K}^{0},K^{*0}\bar{K}^{0}$, and its effects
 are shown in Fig.\ref{li23pli21p}. The constrained $|\lambda'_{i23}
 \lambda'^*_{i21}|$-$\phi_{\spur{R_p}}$ plane is the same
 as the last  plot in Fig.\ref{bounds}.
  $\mathcal{B}(B\to K^{*+}\bar{K}^{0},K^{*0}\bar{K}^{0})$
  are increasing  with $|\lambda'_{i23} \lambda'^*_{i21}|$,
  and unaffected by $\phi_{\spur{R_p}}$.
   $\mathcal{A}^{dir}_{CP}(B\to K^{0}\bar{K}^{0})$
  is steady against  $|\lambda'_{i23} \lambda'^*_{i21}|$, and
   $|\mathcal{A}^{dir}_{CP}(B\to K^{*+}\bar{K}^{0},K^{*0}\bar{K}^{0})|$
  could be varied in small  ranges with $|\lambda'_{i23} \lambda'^*_{i21}|$.
  $\mathcal{A}^{dir}_{CP}(B\to K^{0}\bar{K}^{0})$ is decreasing
  with $|\phi_{\spur{R_p}}|$, but $\mathcal{A}^{dir}_{CP}(B\to K^{*+}\bar{K}^{0},K^{*0}
   \bar{K}^{0})$  are increasing with $|\phi_{\spur{R_p}}|$.

\end{itemize}

The predictions of $\mathcal{B}$ and $\mathcal{A}^{dir}_{CP}$ are
quite uncertain in the RPV SUSY, since we just have few experimental
measurements  and many theoretical uncertainties. One must wait for
the error bars to come down and more channels measured. With the
operation of $B$ factory experiments, large amounts of experimental
data on hadronic $B$ meson decays are being collected, and
measurements of previously known observable will become more
precise. From the comparison of our predictions in
Figs.\ref{li23ppli12pp}-\ref{li23pli21p} with the near future
experiments,   one will obtain more stringent bounds on the product
combinations of RPV couplings. On the other hand, the RPV SUSY
predictions of other decays will become more precise by the more
stringent bounds on the  RPV couplings.

\section{Conclusions}
In conclusions, the pure penguin $B \to K^{(*)}\bar{K}^{(*)}$ decays
are very important for understanding the dynamics of nonleptonic
two-body $B$ decays and testing the SM. We have studied the $B\to
K^{(*)}\bar{K}^{(*)}$ decays with the QCDF approach in the RPV SUSY
model. We have obtained  fairly constrained parameter spaces of the
RPV couplings from the present experimental data of $B \to
K^{(*)}\bar{K}^{(*)}$ decays,  and some of these constraints are
stronger than the existing ones. Furthermore, using the constrained
parameter spaces, we have shown the RPV SUSY expectations for the
other quantities in $B \to K^{(*)}\bar{K}^{(*)}$ decays which have
not been measured yet. We have found that the RPV effects could
significantly
 alter   $\mathcal{B}$ and
$\mathcal{A}^{dir}_{CP}$ from their SM values, but $f_L(B\rightarrow
K^{*+}\bar{K}^{*0},K^{*0}\bar{K}^{*0})$ are not significantly
affected by the RPV effects and the decays $B\rightarrow
K^{*+}\bar{K}^{*0},K^{*0}\bar{K}^{*0}$ are still dominated by the
longitudinal polarization.  We also have presented correlations
between the physical observable $\mathcal{B}$,
$\mathcal{A}^{dir}_{CP}$, $f_L$ and the constrained  parameter
spaces of RPV couplings in Figs.\ref{li23ppli12pp}-\ref{li23pli21p},
which could be tested  in the near future.

\section*{Acknowledgments}
 The work is supported  by National Science
Foundation under contract No.10305003, Henan Provincial Foundation
for Prominent Young Scientists under contract No.0312001700 and the
NCET Program  sponsored by Ministry of Education, China.

\begin{appendix}
 \begin{center}
 {\LARGE{\bf Appendix}}
 \end{center}
\section{\hspace{-0.6cm}. Correction functions for
 $B\to M_1M_2$ decay at $\alpha_s$ order}
In this appendix, we present the explicit form for the correction
functions appearing in the parameters $a^p_i$ and $b^p_i$. It's
noted that in $B\to PV$ decays, $\Phi_{M}(u)\rightarrow
\Phi^{V}_{\parallel}(u)$ if M a vector meson.

\subsection{The correction functions in $B\to PP,PV$ decays }

\hspace{0.6cm}{\Large $\bullet$} One-loop vertex correction function
is
\begin{eqnarray}
 V_{M_2}=12\ln \frac{m_b}{\mu}-18+3 \int_0^1{\rm d}u
 \left(\frac{1-2u}{1-u}\ln u-i \pi\right)\Phi_{M_2}(u).
\end{eqnarray}

{\Large $\bullet$} The hard spectator interactions are given by
\begin{eqnarray}
H_{M_1M_2}=\frac{4\pi^2}{N_C}\frac{f_B f_{M_1}}{m_B^2 F_0^{B \to
M_1}(m^2_{M_2})}\int^1_0\frac{{\rm
d}\xi}{\xi}\Phi^B_1(\xi)\int^1_0\frac{{\rm d}u
}{\bar{u}}\Phi_{M_2}(u)\int^1_0\frac{{\rm d}v
}{\bar{v}}\left[\Phi_{M_1}(v)+\frac{2\mu_{M_1}}{M_B}\Phi^{M_1}_p(v)\right],
\end{eqnarray}
if $M_2$ is a pseudoscalar meson, and
\begin{eqnarray}
H_{M_1M_2}=\frac{4\pi^2}{N_C}\frac{f_B f_{M_1}}{m_B^2 A_0^{B \to
M_1}(m^2_{M_2})}\int^1_0\frac{{\rm
d}\xi}{\xi}\Phi^B_1(\xi)\int^1_0\frac{{\rm d}u
}{\bar{u}}\Phi_{M_2}(u)\int^1_0\frac{{\rm d}v
}{\bar{v}}\Phi_{M_1}(v),
\end{eqnarray}
if $M_2$ is a vector meson.

Considering the off-shellness of the gluon in hard scattering
kernel, it is natural to associate a scale $\mu_h \sim
\sqrt{\Lambda_{QCD} m_b}$ , rather than $\mu\sim m_b$. For the
logarithmically divergent integral,
 we will parameterize it as in \cite{benekeNPB675}:
  $X_H=\int^1_0 {\rm d}u/u=-ln (\Lambda_{QCD}/m_b)+\varrho_He^{i\phi_H}$
$m_b/\Lambda_{QCD}$  with $(\varrho_H,\phi_H)$ related to the
contributions from hard spectator scattering. In the numerical
analysis, we  take $\Lambda_{QCD}=0.5GeV$,
$(\varrho_h,\phi_H)=(0,0)$ as our default values. The same as in $B
\to VV$ decay.

 {\Large $\bullet$} The penguin contributions at  the twist-2
are described by the functions
\begin{eqnarray}
P^p_{M_2,2}&=&C_1G_{M_2}(s_p)+C_3\Big[G_{M_2}(0)+G_{M_2}(1)\Big]
+(C_4+C_6)\bigg[(n_f-2)G_{M_2}(0)\nonumber\\
&&+~G_{M_2}(s_c)
+G_{M_2}(1)-\frac{2n_f}{3}\bigg]
-C^{eff}_{8g}\int_0^1 {\rm d}u ~\frac{2\Phi_{M_2} (u)}{1-u},\nonumber\\
P^{p,EW}_{M_2,2}&=&
(C_1+N_CC_2)G_{M_2}(s_p)-~C^{eff}_{7\gamma}\int_0^1 {\rm d}u
~\frac{3\Phi_{M_2} (u)}{1-u},
\end{eqnarray}
where $n_f = 5$ is the number of quark flavors, and $s_u = 0, s_c =
(m_c/m_b)^2$ are mass ratios involved in the evaluation of the
penguin diagrams. The function $G_{M_2}(s)$ is defined as
 \begin{eqnarray}
G_{M_2}(s)=\frac{2}{3}+\frac{4}{3}\ln \frac{m_b}{\mu}+4\int_0^1{\rm
d}u \int_0^1{\rm d}x~ x\bar{x}\ln{(s-x\bar{x}\bar{u} -i
\epsilon)}\Phi_{M_2} (u).
 \end{eqnarray}

{\Large $\bullet$} The twist-3 terms from the penguin diagrams are
given by
\begin{eqnarray}
P^p_{M_2,3}&=&C_1\widehat{G}_{M_2}(s_p)+C_3\left[\widehat{G}_{M_2}(0)
+\widehat{G}_{M_2}(1)\right]+~(C_4+C_6)\bigg[(n_f-2)\widehat{G}_{M_2}(0)
\nonumber\\
&&+~\widehat{G}_{M_2}(s_c)+\widehat{G}_{M_2}(1)-\frac{2n_f}{3}\bigg]
-~2 C^{eff}_{8g},\nonumber\\
P^{p,EW}_{M_2,3}&=& (C_1+N_CC_2)\widehat{G}_{M_2}(s_p)-~3
C^{eff}_{7\gamma},
\end{eqnarray}
with
\begin{eqnarray}
\widehat{G}_{M_2}(s)=\frac{2}{3}+\frac{4}{3}\ln
\frac{m_b}{\mu}+4\int_0^1{\rm d}u \int_0^1{\rm d}x~
x\bar{x}\ln{(s-x\bar{x}\bar{u} -i \epsilon)}\Phi^{M_2}_p (u),
\end{eqnarray}
if $M_2$ is a pseudoscalar meson, and we omit the  twist-3 terms
from the penguin diagrams when $M_2$ is  a vector  meson.

{\Large $\bullet$} The weak annihilation contributions are given by
\begin{eqnarray}
&&A^i_1(M_1,M_2)\approx A^i_2(M_1,M_2) \approx \pi \alpha_s \left[18
\left( X_A-4+\frac{\pi^2}{3}\right)+ 2r^{M_1}_\chi r^{M_2}_\chi
X^2_A
\right],\nonumber \\
&&A^i_3(M_1,M_2)\approx6\pi\alpha_s (r^{M_1}_\chi-r^{M_2}_\chi)
\left(
X^2_A-2X_A+\frac{\pi^2}{3}\right),\nonumber\\
&&A^f_3(M_1,M_2)\approx 6 \pi \alpha_s
(r^{M_1}_\chi+r^{M_2}_\chi)\left(
2X^2_A-X_A\right), \nonumber \\
&&A^f_1(M_1,M_2)=0, ~~~~~ A^f_2(M_1,M_2)=0.
\label{An1}
\end{eqnarray}
when both final state mesons are pseudoscalar, whereas
\begin{eqnarray}
&&A^i_1(M_1,M_2)\approx -A^i_2(M_1,M_2) \approx 18\pi \alpha_s
\left(
X_A-4+\frac{\pi^2}{3}\right),\nonumber \\
&&A^i_3(M_1,M_2)\approx6\pi\alpha_s r^{M_1}_\chi \left(
X^2_A-2X_A+\frac{\pi^2}{3}\right),\nonumber\\
&&A^f_3(M_1,M_2)\approx- 6 \pi \alpha_s r^{M_1}_\chi\left(
2X^2_A-X_A\right), \nonumber \\
&&A^f_1(M_1,M_2)=0, ~~~~~ A^f_2(M_1,M_2)=0.
\label{An2}
\end{eqnarray}
when $M_1$ is a vector meson and $M_2$ is a pseudoscalar. For the
opposite case of a pseudoscalar $M_1$ and a vector $M_2$, one
exchanges $r^{M_1}_\chi\leftrightarrow r^{M_2}_\chi$ in the previous
equations and changes the sign of $A^f_3$.

 Here the superscripts $i$ and $f$ refer to gluon emission from the initial
and final state quarks, respectively. The subscript $k$ of
$A^{i,f}_k$ refers to one of the three possible Dirac structures
$\Gamma_1 \otimes \Gamma_2$, namely $k = 1$ for $(V-A)\otimes(V-A)$,
$k = 2$ for $(V-A)\otimes(V+A)$, and $k = 3$ for
$(-2)(S-P)\otimes(S+P)$.  $X_A = \int^1_0 {\rm d}u/u$ is a
logarithmically divergent integral, and will be phenomenologically
parameterized in the calculation as $X_H$. As for the hard spectator
terms, we will evaluate the various quantities in Eqs. (\ref{An1})
and (\ref{An2}) at the scale $\mu_h = \sqrt{\Lambda_{QCD} m_b}$.

\subsection{$B\to VV$ decays}
In the rest frame of $B$ system, since the $B$ meson has spin zero,
two vectors have the same helicity therefore three polarization
states are possible, one longitudinal (L) and two transverse,
corresponding to helicities $\lambda=0$ and $\lambda=\pm$ ( here
$\lambda_1=\lambda_2=\lambda$). We assume the $M_1$($M_2$) meson
flying in the minus(plus) z-direction carrying the momentum
$p_1$($p_2$), Using the sign convention $\epsilon^{0123}=-1$, we
have {\footnotesize
\begin{eqnarray}
 A_{M_1M_2}=\left \{\begin{array}{ll}\frac{i
f_{M_2}}{2m_{M_1}}\left[(m^2_{B}-m_{M_1}^2-m_{M_2}^2)(m_B+m_{M_1})
A_1^{B\to M_1}(m_{M_2}^2)-\frac{4m_{B}^2p_c^2}{m_{B}+m_{M_1}}
A_2^{B\to M_1}(m_{M_2}^2)\right]\equiv h_0,
\\i
 f_{V_2}m_{M_2}[(m_{B}+m_{M_1})A_1 ^{B\to M_1}(m_{M_2}^2)\mp
\frac{2m_{B}p_c}{m_{B}+m_{M_1}}V ^{B\to M_1}(m_{M_2}^2)]\equiv
h_{\pm},
 \end{array}
 \right.
 \end{eqnarray}}
 where $h_0$ for $\lambda=0$ and $h_\pm$ for
$\lambda=\pm$.

{\Large $\bullet$} $V_{M_2}^{\lambda}(\pm 1)$ contain the
contributions from the vertex corrections, and given by
 \begin{eqnarray}
 &&V_{M_2}^0(a)=12\ln \frac{m_b}{\mu}-18+\int_0^1{\rm d}u
\Phi_{\parallel}^{M_2} (u)\left(3\frac{1-2u}{1-u}\ln u-3i \pi\right),\\
 &&V_{M_2}^{\pm}(a)=12\ln \frac{m_b}{\mu}-18+\int_0^1{\rm d}u
\left(\mbox{g}_{\perp}^{(v){M_2}} (u)\pm \frac{a
\mbox{g}_{\perp}^{\prime(a){M_2}}
(u)}{4}\right)\left(3\frac{1-2u}{1-u}\ln u-3i \pi\right).\nonumber
 \end{eqnarray}

{\Large $\bullet$} The hard spectator scattering contributions,
explicit calculations for $H_{M_1M_2}^\lambda (a)$ yield
\begin{eqnarray}
H_{M_1M_2}^0(a)&=&\frac{4\pi^2}{N_C}\frac{if_{B}
f_{V_1}f_{V_2}}{h_0}\int_0^1{\rm d}\xi \frac{\Phi_1^B
(\xi)}{\xi}\int_0^1{\rm d}v \frac{\Phi_{\parallel}^{M_1}
(v)}{\bar{v}}\int_0^1{\rm d}u \frac{\Phi_{\parallel}^{M_2}
(u)}{u},\nonumber\\
H_{M_1M_2}^{\pm}(a)&=&-\frac{4
\pi^2}{N_C}\frac{2if_{B}f^{\perp}_{M_1}f_{M_2}m_{M_2}}{m_{B}h_{\pm}}
(1\mp1)\int_0^1{\rm d}\xi \frac{\Phi_1^B (\xi)}{\xi}\int_0^1{\rm d}v
\frac{\Phi_{\perp}^{M_1} (v)}{\bar{v}^2}\nonumber
\\&&\times\int_0^1{\rm d}u \left(\mbox{g}_{\perp}^{(v){M_2}}
(u)-\frac{a\mbox{g}_{\perp}^{\prime(a){M_2}} (u)}{4}\right)+\frac{4
\pi^2}{N_C}\frac{2if_{B}f_{M_1}f_{M_2}m_{M_1}m_{M_2}}{m_{B}^2h_{\pm}}
\int_0^1{\rm d}\xi \frac{\Phi_1^B (\xi)}{\xi}\nonumber
\\&&\times\int_0^1{\rm d}v {\rm d}u\left(\mbox{g}_{\perp}^{(v){M_1}} (v)
\pm\frac{\mbox{g}_{\perp}^{\prime(a){M_1}}
(v)}{4}\right)\left(\mbox{g}_{\perp}^{(v){M_2}} (u)\pm \frac{a
\mbox{g}_{\perp}^{\prime(a){M_2}} (u)}{4}\right)\frac{u+\bar v}
{u{\bar v}^2},
\end{eqnarray}
with $\bar{v}=1-v$,  when the asymptotical form for the vector meson
LCDAs adopted, there will be infrared divergences  in
$H_{M_1M_2}^{\pm}$. As in \cite{benekeNPB606,ykc}, we introduce a
cutoff of order $\Lambda_{QCD}/m_b$ and take $\Lambda_{QCD}=0.5$ GeV
as our default value.

{\Large $\bullet$} The contributions of the QCD penguin-type
diagrams can be described by the functions
\begin{eqnarray}
P^{\lambda
,p}_{M_2,2}&=&C_1G^{\lambda}_{M_2}(s_p)+C_3\left[G^{\lambda}_{M_2}
(s_q)+G^{\lambda}_{M_2}(s_b)\right]
+(C_4+C_6)\sum_{q'=u}^b \Big[G^\lambda
_{M_2}(s_{q'})-\frac{2}{3}\Big]
\nonumber\\
&&+\frac{3}{2}C_9\Big[e_qG^\lambda
_{M_2}(s_q)+e_bG^\lambda_{M_2}(s_b)\Big]+ \frac{3}{2}(C_8+C_{10})
\sum_{q'=u}^be_{q'}\Big[G^\lambda_{M_2}(s_{q'})-\frac{2}{3}\Big]+C^{eff}_{8g}
G_g^\lambda ,\nonumber\\
P^{\lambda,p,EW}_{M_2,2}&=&
(C_1+N_CC_2)\left[\frac{2}{3}+\frac{4}{3}\ln
\frac{m_b}{\mu}-G^{\lambda}_{M_2}(s_p)\right]+~\frac{3}{2}
C^{eff}_{7\gamma}G^{\lambda}_g,
\end{eqnarray}
\begin{eqnarray}
G^0_{M_2} (s)&=&\frac{2}{3}+\frac{4}{3}\ln
\frac{m_b}{\mu}+4\int_0^1{\rm d}u ~\Phi_{\parallel}^{M_2}
(u) \mbox{g}(u,s),\nonumber\\
G^{\pm}_{M_2}(s)&=&\frac{2}{3}+\frac{2}{3}\ln
\frac{m_b}{\mu}+2\int_0^1{\rm d}u
~(\mbox{g}_{\perp}^{(v){M_2}}(u)\pm
\frac{\mbox{g}_{\perp}^{\prime(a){M_2}}(u)}{4}) \mbox{g}(u,s),
\end{eqnarray}
with the function g$(u,s)$ defined as
 \begin{eqnarray}
 \mbox{g}(u,s)&=&\int_0^1{\rm d}x~ x\bar{x}\ln{(s-x\bar{x}\bar{u} -i \epsilon)}.
 \end{eqnarray}
 We omit the  twist-3 terms
from the penguin diagrams for $B \to VV$ decays.

{\Large $\bullet$} We have also taken into account the contributions
of the dipole operator $O_{8g}$, which are described by the
functions
\begin{eqnarray}
&&G_g^0=-\int_0^1 {\rm d}u ~\frac{2\Phi_{\parallel}^{M_2}
(u)}{1-u},\nonumber\\
&&G_g^{\pm}=\int_0^1 ~\frac{{\rm d}u}{\bar{u}}\left[
-\bar{u}\mbox{g}_{\perp}^{(v){M_2}}(u)\mp
\frac{\bar{u}\mbox{g}_{\perp}^{\prime(a){M_2}}(u)}{4} +\int_0^u{\rm
d}v\left(\Phi_\parallel^{M_2}(v)-\mbox{g}_\perp^{(v)M_2}(v)\right)
+\frac{\mbox{g}_{\perp}^{(a){M_2}}(u)}{4}\right]\label{o8g},
\end{eqnarray}
here we consider the higher-twist effects $k^{\mu}
=uEn_-^\mu+k_\perp^\mu+\frac{\vec{k}_\perp^2}{4uE}n_+^\mu$ in the
projector of the vector meson.  The $G_g^{\pm}=0$ in Eq.(\ref{o8g})
 \cite{ykc,kagan} if considering  the Wandzura-Wilczek-type relations \cite{WWR}.

We have not onsidered the annihilation contributions in $B \to VV$
decays.
\subsection{The contributions of new operators in RPV SUSY}
Compared  with the operators in the $\mathcal{H}^{SM}_{eff}$, there
are new operators $(\bar{q}_2q_3)_{V\pm A} (\bar{b}q_1)_{V+A}$ in
the $\mathcal{H}^{\spur{R_p}}_{eff}$.

{\Large $\bullet$} For $B \to PP,PV$ decays, since
\begin{eqnarray}
&&\langle P|~ \bar{q}_1\gamma_\mu(1-\gamma_{5})q_2|~ 0
\rangle=-\langle P|~ \bar{q}_1\gamma_\mu(1+\gamma_{5})q_2|~ 0
\rangle=-\langle P|~ \bar{q}_1 \gamma_\mu\gamma_{5}q_2|~ 0 \rangle
,\nonumber\\
&&\langle P|~ \bar{q}\gamma_{\mu}(1-\gamma_{5})b~|~B \rangle
=\langle P|~ \bar{q}\gamma_{\mu}(1+\gamma_{5})b~|~B \rangle=\langle
P|~ \bar{q}\gamma_{\mu}b~|~B \rangle,\hspace{1cm} \nonumber\\
&&\langle V|~ \bar{q}_1\gamma_\mu(1-\gamma_{5})q_2|~ 0
\rangle=\langle V|~ \bar{q}_1\gamma_\mu(1+\gamma_{5})q_2|~ 0
\rangle=\langle V|~ \bar{q}_1 \gamma_\mu q_2|~ 0 \rangle
,\nonumber\\
&&\langle V|~ \bar{q}\gamma_{\mu}(1-\gamma_{5})b~|~B \rangle
=-\langle V|~ \bar{q}\gamma_{\mu}(1+\gamma_{5})b~|~B
\rangle=-\langle V|~ \bar{q}\gamma_{\mu}\gamma_{5}b~|~B
\rangle,\hspace{1cm}
\end{eqnarray}
the RPV contribution to the decay amplitude will  modify the SM
amplitude by an overall  relation.

{\Large $\bullet$} For $B \to VV$, we will use the prime on the
quantities stands for  the $(\bar{q}_2q_3)_{V\pm A}
(\bar{b}q_1)_{V+A}$ current contribution. In the NF approach, the
factorizable amplitude can be expressed as
\begin{eqnarray}
  A'_{M_1M_2}&=&\langle M_2|(\bar{q}_2\gamma_\mu(1-a\gamma_5)q_3)|0\rangle
 \langle M_1|(\bar{b}\gamma^\mu(1+\gamma_5)q_1)|B\rangle.
 \end{eqnarray}

Taking the $M_1$($M_2$) meson flying  in the minus(plus) z-direction
and using the sign convention $\epsilon^{0123}=-1$, we have
{\footnotesize
\begin{eqnarray}
 A'_{M_1M_2}=\left \{\begin{array}{ll}\frac{-i
f_{M_2}}{2m_{M_1}}\left[(m^2_{B}-m_{M_1}^2-m_{M_2}^2)(m_B+m_{M_1})
A_1^{B\to M_1}(m_{M_2}^2)-\frac{2m_{B}^2p_c^2}{m_{B}+m_{M_1}} A_2^{B
\to M_1}(m_{M_2}^2)\right]\equiv h'_0,
\\-i
 f_{M_2}m_{M_2}\left[(m_{B}+m_{M_1})A_1 ^{B\to M_1}(m_{M_2}^2)\pm
\frac{2m_{B}p_c}{m_{B}+m_{M_1}}V ^{B\to M_1}(m_{M_2}^2)\right]\equiv
h'_{\pm}.
 \end{array}
 \right.
 \end{eqnarray}}

The  vertex corrections $V^{'\lambda}_{M_2}(a)$ and the hard
spectator scattering corrections $H^{'\lambda}_{M_1M_2}(a)$ as
follows: {\footnotesize
\begin{eqnarray}
 V_{M_2}^{'0}(a)&=&-12\ln \frac{m_b}{\mu}+18-6(1+a)-\int_0^1{\rm d}u
\Phi_{\parallel}^{M_2} (u)\left(3\frac{1-2u}{1-u}\ln u-3i \pi\right),
\nonumber\\
 V_{M_2}^{'\pm}(a)&=&-12\ln \frac{m_b}{\mu}+18-6(1+a)-\int_0^1{\rm d}u
\left(g_{\perp}^{(v){M_2}} (u)\pm \frac{a g_{\perp}^{\prime(a){M_2}}
(u)}{4}\right)\left(3\frac{1-2u}{1-u}\ln u-3i \pi\right),\nonumber\\
H^{'0}_{M_1M_2}(a)&=&\frac{4\pi^2}{N_C}\frac{if_{B}f_{M_1}
f_{M_2}}{h^{'}_0}\int_0^1{\rm
d}\xi \frac{\Phi_1^B (\xi)}{\xi}\int_0^1{\rm d}v
\frac{\Phi_{\parallel}^{M_1} (v)}{\bar{v}}\int_0^1{\rm d}u
\frac{\Phi_{\parallel}^{M_2}
(u)}{u},\nonumber\\
H_{M_1M_2}^{'\pm}(a)&=&-\frac{4
\pi^2}{N_C}\frac{2if_{B}f^{\perp}_{M_1}f_{M_2}m_{M_2}}{m_{B}h^{'}_{\pm}}
(1\pm1)\int_0^1{\rm d}\xi \frac{\Phi_1^B (\xi)}{\xi}\int_0^1{\rm d}v
\frac{\Phi_{\perp}^{M_1}
(v)}{\bar{v}^2}\nonumber\\
&&\times\int_0^1{\rm d}u \left(g_{\perp}^{(v){M_2}}(u)
+\frac{ag_{\perp}^{\prime(a){M_2}} (u)}{4}\right) +\frac{4
\pi^2}{N_C}\frac{2if_{B}f_{M_1}f_{M_2}m_{M_1}m_{M_2}}
{m_{B}^2h^{'}_{\pm}}\int_0^1{\rm
d}\xi \frac{\Phi_1^B (\xi)}{\xi}\nonumber\\
&&\times\int_0^1{\rm d}v {\rm d}u\left(g_{\perp}^{(v){M_1}} (v)
\mp\frac{g_{\perp}^{\prime(a){M_1}}
(v)}{4}\right)\left(g_{\perp}^{(v){M_2}} (u)\pm \frac{a
g_{\perp}^{\prime(a){M_2}} (u)}{4}\right)\frac{u+\bar v} {u{\bar
v}^2}.
\end{eqnarray}}

\section{\hspace{-0.6cm}. The amplitudes in the SM}\vspace{-0.8cm}
{\scriptsize
\begin{eqnarray}
\mathcal{A}^{SM}_f(B^+ \rightarrow K^+
\bar{K}^{0})&=&\frac{G_F}{\sqrt{2}}\left\{-V^*_{tb}V_{td}
 \left[ a_4-\frac{1}{2}a_{10}+r^{K^0}_{\chi}(a_6-\frac{1}{2}a_{8})
  \right] \right\}
A_{K^{+}\bar{K}^0},\\
\mathcal{A}^{SM}_a(B^+ \rightarrow K^+
\bar{K}^{0})&=&i\frac{G_F}{\sqrt{2}}f_Bf^{2}_{K}\left\{V^*_{ub}
V_{ud}b_2(K^+,\bar{K}^0)-V^*_{tb}V_{td}
 \left[ b_3(K^+,\bar{K}^0)+b^{ew}_3(K^+,\bar{K}^0) \right] \right\},\\
\mathcal{A}^{SM}_f(B^0 \rightarrow K^0
\bar{K}^{0})&=&\frac{G_F}{\sqrt{2}}\left\{-V^*_{tb}V_{td}
 \left[ a_4-\frac{1}{2}a_{10}+r^{K^0}_{\chi}(a_6-\frac{1}{2}a_{8})
 \right] \right\}
A_{K^{0}\bar{K}^0},\\
\mathcal{A}^{SM}_a(B^0 \rightarrow K^0
\bar{K}^{0})&=&i\frac{G_F}{\sqrt{2}}f_Bf^{2}_{K}\left\{-V^*_{tb}V_{td}
 \left[ \frac{}{}b_3(\bar{K}^0,K^0)+b_4(\bar{K}^0,K^0)+b_4(K^0,\bar{K}^0)
  \right.\right.\nonumber\\
 &&\left.\left.
 -\frac{1}{2}b^{ew}_3(\bar{K}^0,K^0)-\frac{1}{2}b^{ew}_4(\bar{K}^0,K^0)
 -\frac{1}{2}b^{ew}_4(K^0,\bar{K}^0)\right] \right\},\\
\mathcal{A}^{SM}_f(B^+ \rightarrow
K^{*+}\bar{K}^0)&=&\frac{G_F}{\sqrt{2}}\left\{-V^*_{tb}V_{td}
 \left[ a_4-\frac{1}{2}a_{10}-r^{K^0}_{\chi}(a_6-\frac{1}{2}a_{8})
  \right] \right\}A_{K^{*+}\bar{K}^0},\\
\mathcal{A}^{SM}_a(B^+ \rightarrow
K^{*+}\bar{K}^0)&=&\frac{G_F}{\sqrt{2}}f_Bf_Kf_{K^*}
\left\{V^*_{ub}V_{ud}b_2(K^{*+},\bar{K}^{0})-V^*_{tb}V_{td}
 \left[ b_3(K^{*+},\bar{K}^{0})+b^{ew}_3(K^{*+},\bar{K}^{0})
  \right] \right\},\\
\mathcal{A}^{SM}_f(B^+ \rightarrow K^+
\bar{K}^{*0})&=&\frac{G_F}{\sqrt{2}}\left\{-V^*_{tb}V_{td}
 \left[ a_4-\frac{1}{2}a_{10} \right] \right\}
A_{K^{+}\bar{K}^{*0}},\\
\mathcal{A}^{SM}_a(B^+ \rightarrow K^+
\bar{K}^{*0})&=&\frac{G_F}{\sqrt{2}}f_Bf_Kf_{K^*}
\left\{V^*_{ub}V_{ud}b_2(K^{+},\bar{K}^{*0})-V^*_{tb}V_{td}
 \left[ b_3(K^{+},\bar{K}^{*0})+b^{ew}_3(K^{+},\bar{K}^{*0})
 \right] \right\},\\
\mathcal{A}^{SM}_f(B^0 \rightarrow
K^{*0}\bar{K}^0)&=&\frac{G_F}{\sqrt{2}}\left\{-V^*_{tb}V_{td}
 \left[ a_4-\frac{1}{2}a_{10}-r^{K^0}_{\chi}(a_6-\frac{1}{2}a_{8})
  \right] \right\}
A_{K^{*0}\bar{K}^0},\\
\mathcal{A}^{SM}_a(B^0 \rightarrow
K^{*0}\bar{K}^0)&=&\frac{G_F}{\sqrt{2}}f_Bf_Kf_{K^*}\left\{-V^*_{tb}V_{td}
 \left[\frac{}{} b_3(K^{*0},\bar{K}^0)+b_4(K^{*0},\bar{K}^0)
 +b_4(\bar{K}^0,K^{*0}) \right.\right.\nonumber\\
 &&\left.\left.
 -\frac{1}{2}b^{ew}_3(K^{*0},\bar{K}^0)-\frac{1}{2}b^{ew}_4(K^{*0},
 \bar{K}^0)-\frac{1}{2}b^{ew}_4(\bar{K}^0,K^{*0})\right] \right\},\\
\mathcal{A}^{SM}_f(B^0 \rightarrow K^0
\bar{K}^{*0})&=&\frac{G_F}{\sqrt{2}}\left\{-V^*_{tb}V_{td}
 \left[ a_4-\frac{1}{2}a_{10} \right] \right\}
A_{K^{0}\bar{K}^{*0}},\\
\mathcal{A}^{SM}_a(B^0 \rightarrow K^0
\bar{K}^{*0})&=&\frac{G_F}{\sqrt{2}}f_Bf_Kf_{K^*}\left\{-V^*_{tb}V_{td}
 \left[\frac{}{} b_3(K^0,\bar{K}^{*0})+b_4(K^0,\bar{K}^{*0})
 +b_4(\bar{K}^{*0},K^0) \right.\right.\nonumber\\
 &&\left.\left.
 -\frac{1}{2}b^{ew}_3(K^0,\bar{K}^{*0})-\frac{1}{2}b^{ew}_4(K^0,\bar{K}^{*0})
 -\frac{1}{2}b^{ew}_4(\bar{K}^{*0},K^0)\right] \right\},\\
\mathcal{A}^{SM}_f(B^+ \rightarrow K^{*+}
\bar{K}^{*0})&=&\frac{G_F}{\sqrt{2}}\left\{-V^*_{tb}V_{td}
 \left[ a_4-\frac{1}{2}a_{10} \right] \right\}
A_{K^{*+}\bar{K}^{*0}},\\
\mathcal{A}^{SM}_f(B^0 \rightarrow K^{*0}
\bar{K}^{*0})&=&\frac{G_F}{\sqrt{2}}\left\{-V^*_{tb}V_{td}
 \left[ a_4-\frac{1}{2}a_{10}\right] \right\}
A_{K^{*0}\bar{K}^{*0}},
 \end{eqnarray}}
Here we have not considered the annihilation contributions in $B \to
VV$ decays.

\section{\hspace{-0.6cm}. The amplitudes for RPV}\vspace{-0.8cm}
{\scriptsize
\begin{eqnarray}
\mathcal{A}^{\spur{R}_p}(B^+ \rightarrow K^{+}
\bar{K}^0)&=&\left\{\frac{\lambda''_{i23}\lambda''^*_{i12}}
{16m^2_{\tilde{u}_i}}\eta^{-4/\beta_0}F_{K^{+} \bar{K}^0} +\left(
\frac{\lambda'_{i13}\lambda'^*_{i22}}{8m^2_{\tilde{\nu}_{Li}}}
-\frac{\lambda'_{i22}\lambda'^*_{i31}}{8m^2_{\tilde{\nu}_{Li}}}
\right)\eta^{-8/\beta_0}L_{K^{+} \bar{K}^0} \right.\nonumber\\
&&\left.+\left(
\frac{\lambda'_{i12}\lambda'^*_{i32}}{8m^2_{\tilde{\nu}_{Li}}}
-\frac{\lambda'_{i23}\lambda'^*_{i21}}{8m^2_{\tilde{\nu}_{Li}}}
\right)\eta^{-8/\beta_0}r^{K^0}_\chi \right\}A_{K^{+} \bar{K}^0},\\
\mathcal{A}^{\spur{R}_p}(B^0 \rightarrow K^{0}
\bar{K}^0)&=&\left\{\frac{\lambda''_{i23}
\lambda''^*_{i12}}{16m^2_{\tilde{u}_i}}\eta^{-4/\beta_0}F_{K^{0}
\bar{K}^0} +\left(
\frac{\lambda'_{i13}\lambda'^*_{i22}}{8m^2_{\tilde{\nu}_{Li}}}
-\frac{\lambda'_{i22}\lambda'^*_{i31}}{8m^2_{\tilde{\nu}_{Li}}}
\right)\eta^{-8/\beta_0}L_{K^{0} \bar{K}^0} \right.\nonumber\\
&&\left.+\left(
\frac{\lambda'_{i12}\lambda'^*_{i32}}{8m^2_{\tilde{\nu}_{Li}}}
-\frac{\lambda'_{i23}\lambda'^*_{i21}}{8m^2_{\tilde{\nu}_{Li}}}
\right)\eta^{-8/\beta_0}r^{K^0}_\chi \right\}A_{K^{0} \bar{K}^0},\\
\mathcal{A}^{\spur{R}_p}(B^+ \rightarrow K^{*+}
\bar{K}^0)&=&\left\{\frac{\lambda''_{i23}\lambda''^*_{i12}}
{16m^2_{\tilde{u}_i}}\eta^{-4/\beta_0}F_{K^{*+} \bar{K}^0} +\left(
\frac{\lambda'_{i13}\lambda'^*_{i22}}{8m^2_{\tilde{\nu}_{Li}}}
+\frac{\lambda'_{i22}\lambda'^*_{i31}}{8m^2_{\tilde{\nu}_{Li}}}
\right)\eta^{-8/\beta_0}(-L_{K^{*+} \bar{K}^0}) \right.\nonumber\\
&&\left.-\left(
\frac{\lambda'_{i12}\lambda'^*_{i32}}{8m^2_{\tilde{\nu}_{Li}}}
+\frac{\lambda'_{i23}\lambda'^*_{i21}}{8m^2_{\tilde{\nu}_{Li}}}
\right)\eta^{-8/\beta_0}r^{K^0}_\chi \right\}A_{K^{*+} \bar{K}^0},\\
\mathcal{A}^{\spur{R}_p}(B^+ \rightarrow K^+
\bar{K}^{*0})&=&\left\{\frac{\lambda''_{i23}\lambda''^*_{i12}}
{16m^2_{\tilde{u}_i}}\eta^{-4/\beta_0}F_{K^{+} \bar{K}^{*0}} +\left(
\frac{\lambda'_{i13}\lambda'^*_{i22}}{8m^2_{\tilde{\nu}_{Li}}}
+\frac{\lambda'_{i22}\lambda'^*_{i31}}{8m^2_{\tilde{\nu}_{Li}}}
\right)\eta^{-8/\beta_0}L_{K^{+} \bar{K}^{*0}} \right\}A_{K^{+}
 \bar{K}^{*0}},\\
\mathcal{A}^{\spur{R}_p}(B^0 \rightarrow K^{*0}
\bar{K}^0)&=&\left\{\frac{\lambda''_{i23}\lambda''^*_{i12}}
{16m^2_{\tilde{u}_i}}\eta^{-4/\beta_0}F_{K^{*0} \bar{K}^0} +\left(
\frac{\lambda'_{i13}\lambda'^*_{i22}}{8m^2_{\tilde{\nu}_{Li}}}
+\frac{\lambda'_{i22}\lambda'^*_{i31}}{8m^2_{\tilde{\nu}_{Li}}}
\right)\eta^{-8/\beta_0}(-L_{K^{*0} \bar{K}^0}) \right.\nonumber\\
&&\left.-\left(
\frac{\lambda'_{i12}\lambda'^*_{i32}}{8m^2_{\tilde{\nu}_{Li}}}
+\frac{\lambda'_{i23}\lambda'^*_{i21}}{8m^2_{\tilde{\nu}_{Li}}}
\right)\eta^{-8/\beta_0}r^{K^0}_\chi \right\}A_{K^{*0} \bar{K}^0},\\
\mathcal{A}^{\spur{R}_p}(B^0 \rightarrow K^0
\bar{K}^{*0})&=&\left\{\frac{\lambda''_{i23}\lambda''^*_{i12}}
{16m^2_{\tilde{u}_i}}\eta^{-4/\beta_0}F_{K^{0} \bar{K}^{*0}} +\left(
\frac{\lambda'_{i13}\lambda'^*_{i22}}{8m^2_{\tilde{\nu}_{Li}}}
+\frac{\lambda'_{i22}\lambda'^*_{i31}}{8m^2_{\tilde{\nu}_{Li}}}
\right)\eta^{-8/\beta_0}L_{K^{0} \bar{K}^{*0}} \right\}A_{K^{0}
 \bar{K}^{*0}},\\
\mathcal{A}^{\spur{R}_p}(B^+ \rightarrow K^{*+}
\bar{K}^{*0})&=&\left\{\frac{\lambda''_{i23}\lambda''^*_{i12}}
{16m^2_{\tilde{u}_i}}\eta^{-4/\beta_0}F'_{K^{*+} \bar{K}^{*0}}
+\left(
\frac{\lambda'_{i13}\lambda'^*_{i22}}{8m^2_{\tilde{\nu}_{Li}}}
\right)\eta^{-8/\beta_0}L'_{K^{*+} \bar{K}^{*0}} \right\}A'_{K^{*+}
\bar{K}^{*0}}\nonumber\\
&&+\frac{\lambda'_{i22}\lambda'^*_{i31}}{8m^2_{\tilde{\nu}_{Li}}}L_{K^{*+}
\bar{K}^{*0}} A_{K^{*+}
\bar{K}^{*0}},\\
\mathcal{A}^{\spur{R}_p}(B^+ \rightarrow K^{*0}
\bar{K}^{*0})&=&\left\{\frac{\lambda''_{i23}\lambda''^*_{i12}}
{16m^2_{\tilde{u}_i}}\eta^{-4/\beta_0}F'_{K^{*0} \bar{K}^{*0}}
+\left(
\frac{\lambda'_{i13}\lambda'^*_{i22}}{8m^2_{\tilde{\nu}_{Li}}}
\right)\eta^{-8/\beta_0}L'_{K^{*0} \bar{K}^{*0}} \right\}A'_{K^{*0}
\bar{K}^{*0}}\nonumber\\
&&+\frac{\lambda'_{i22}\lambda'^*_{i31}}{8m^2_{\tilde{\nu}_{Li}}}L_{K^{*0}
\bar{K}^{*0}} A_{K^{*0} \bar{K}^{*0}}.
\end{eqnarray}}

In the $\mathcal{A}^{\spur{R}_p}$,  $F^{(')}_{M_1M_2}$ and
$L^{(')}_{M_1M_2}$ are defined as
\begin{eqnarray}
  F_{M_1M_2}&\equiv& 1-\frac{1}{N_C}+\frac{\alpha_s}{4\pi}
  \frac{C_F}{N_C}\Big[V_{M_2}+H_{M_1M_2}\Big],\\
L_{M_1M_2}&\equiv&\frac{1}{N_C}\left\{
1-\frac{\alpha_s}{4\pi}\frac{C_F}{N_C}\Big[12+V_{M_2}
+H_{M_1M_2}\Big]\right\},
\end{eqnarray}
for $B\rightarrow PP,PV$ decays, and
\begin{eqnarray}
F'_{M_1M_2}&\equiv& 1-\frac{1}{N_C}-\frac{\alpha_s}{4\pi}
\frac{C_F}{N_C}\left[V'^{\lambda}_{M_2}(-1)
+H'^{\lambda}_{M_1M_2}(-1)\right],\\
L'_{M_1M_2}&\equiv&\frac{1}{N_C}\left\{
1+\frac{\alpha_s}{4\pi}\frac{C_F}{N_C}\left[-12
+V'^{\lambda}_{M_2}(1)+H'^{\lambda}_{M_1M_2}(1)\right]\right\},\\
L_{M_1M_2}&\equiv&\frac{1}{N_C}\left\{
1-\frac{\alpha_s}{4\pi}\frac{C_F}{N_C}\left[12
+V^{\lambda}_{M_2}(-1)+H^{\lambda}_{M_1M_2}(-1)\right]\right\},
\end{eqnarray}
for $B\rightarrow VV$ decays.
 \end{appendix}

\begin{figure}[ht]
\begin{center}
\includegraphics[scale=0.75]{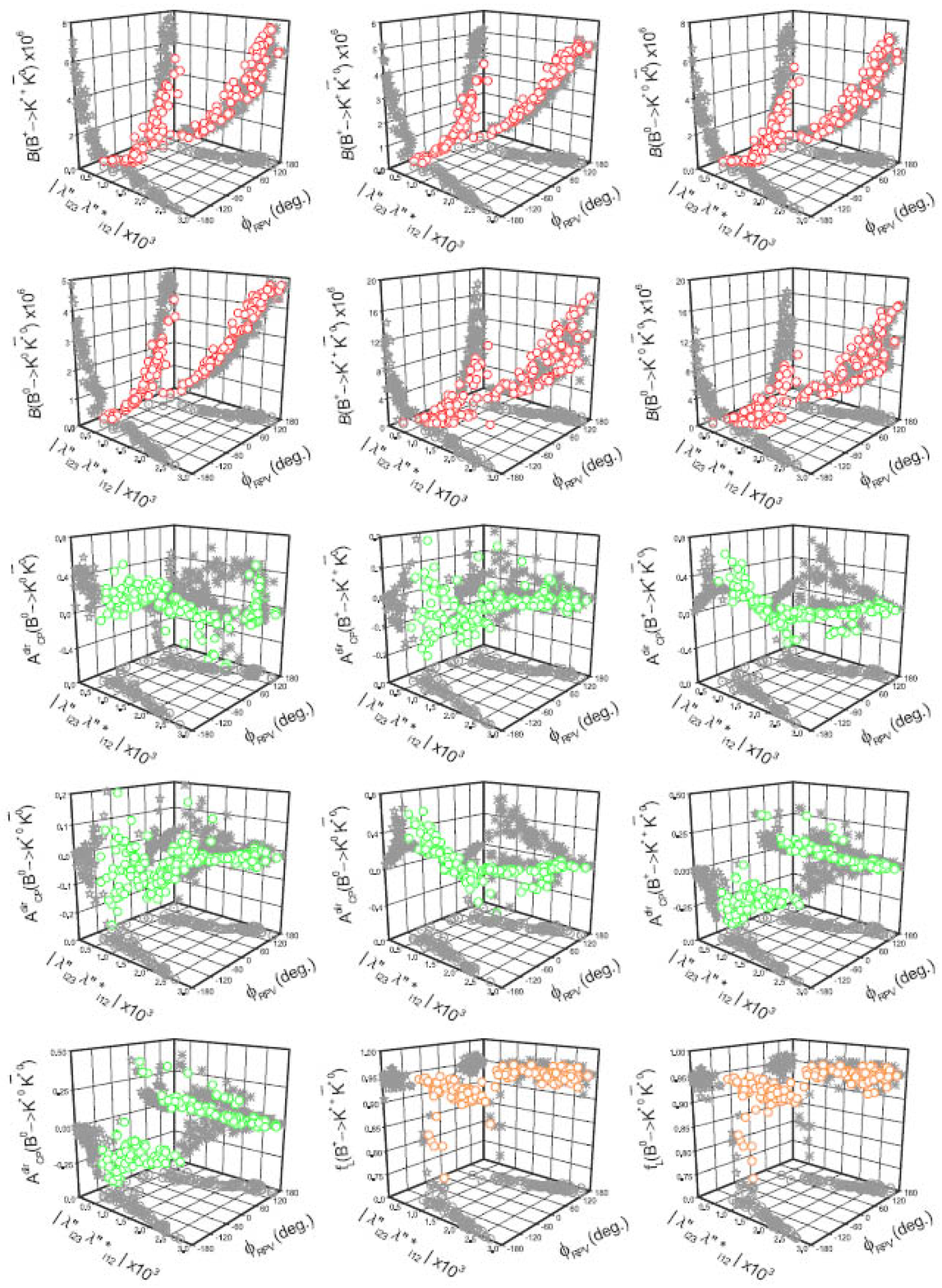}
\end{center}
\vspace{-0.6cm}
 \caption{\small The effects of RPV coupling $\lambda''_{i23}
 \lambda''^*_{i12}$ in $B\to K^{(*)}\bar{K}^{(*)}$ decays.}
 \label{li23ppli12pp}
\end{figure}

\begin{figure}[ht]
\begin{center}
\includegraphics[scale=0.75]{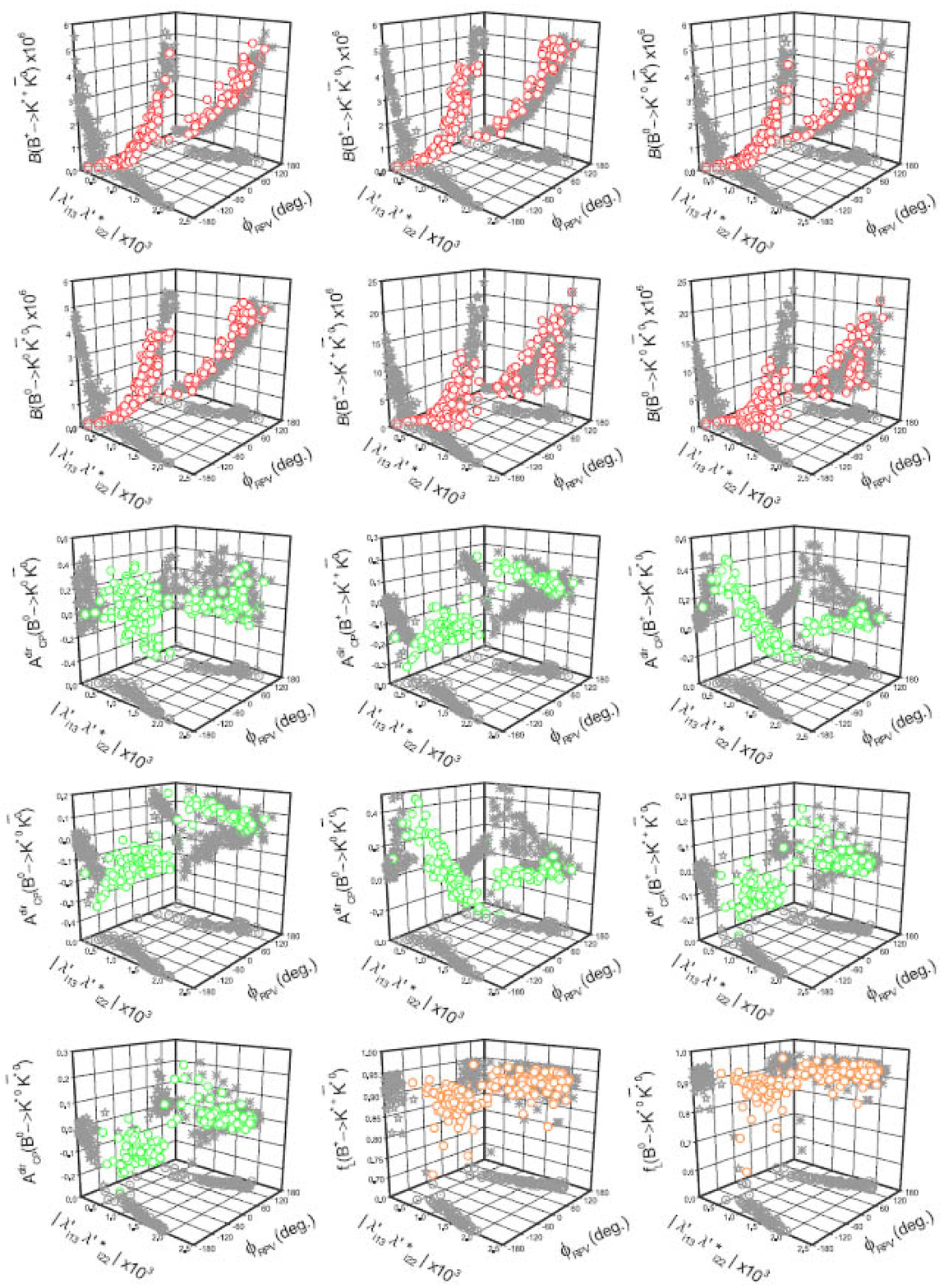}
\end{center}
\vspace{-0.6cm} \caption{\small The effects of RPV coupling
$\lambda'_{i13}
 \lambda'^*_{i22}$ in $B\to K^{(*)}\bar{K}^{(*)}$ decays.}
 \label{li13pli22p}
\end{figure}

\begin{figure}[ht]
\begin{center}
\includegraphics[scale=0.75]{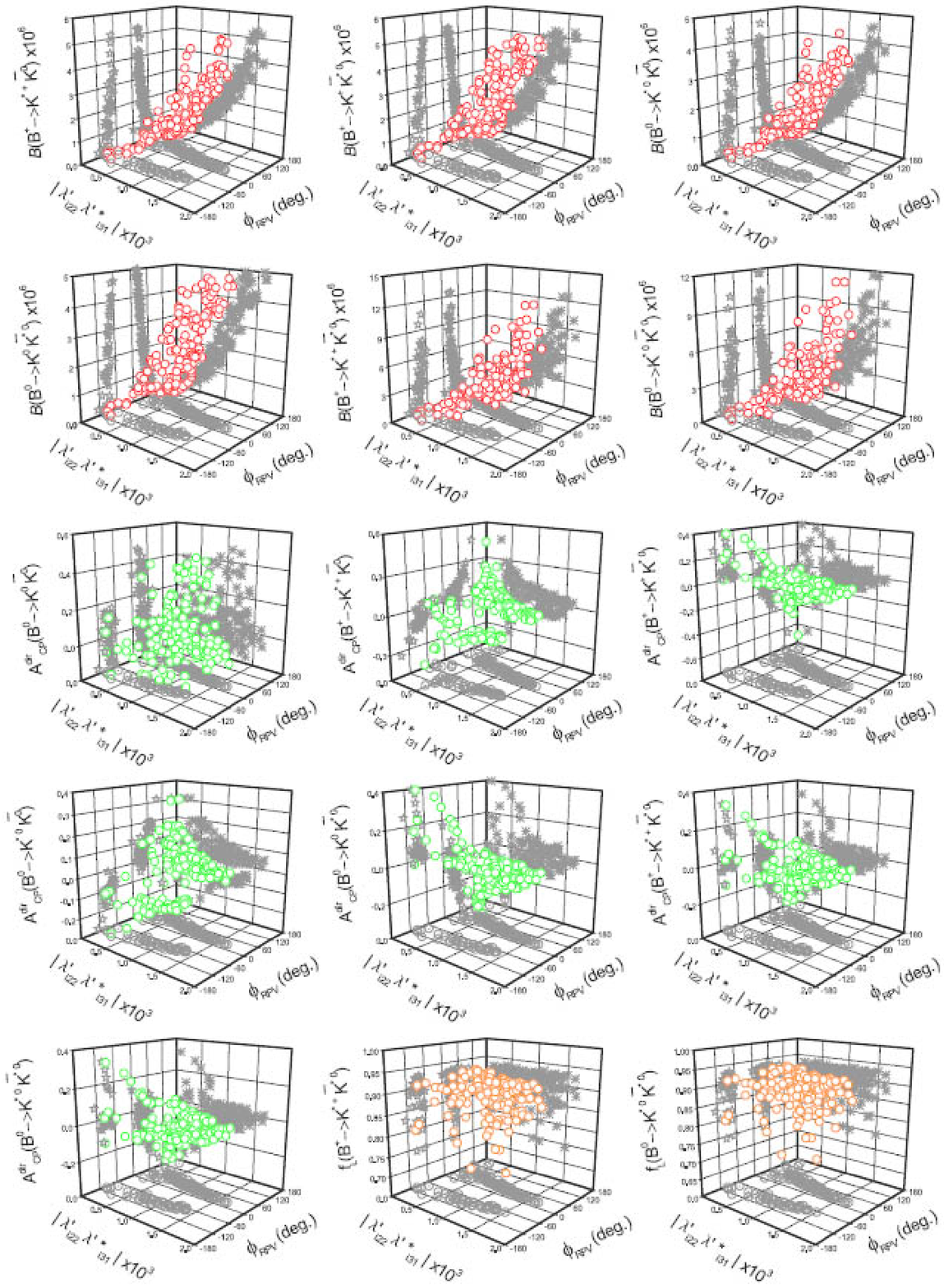}
\end{center}
\vspace{-0.6cm}
 \caption{\small The effects of RPV coupling $\lambda'_{i22}
 \lambda'^*_{i31}$ in $B\to K^{(*)}\bar{K}^{(*)}$ decays.}
 \label{li22pli31p}
\end{figure}

\begin{figure}[htbp]
\begin{center}
\includegraphics[scale=0.75]{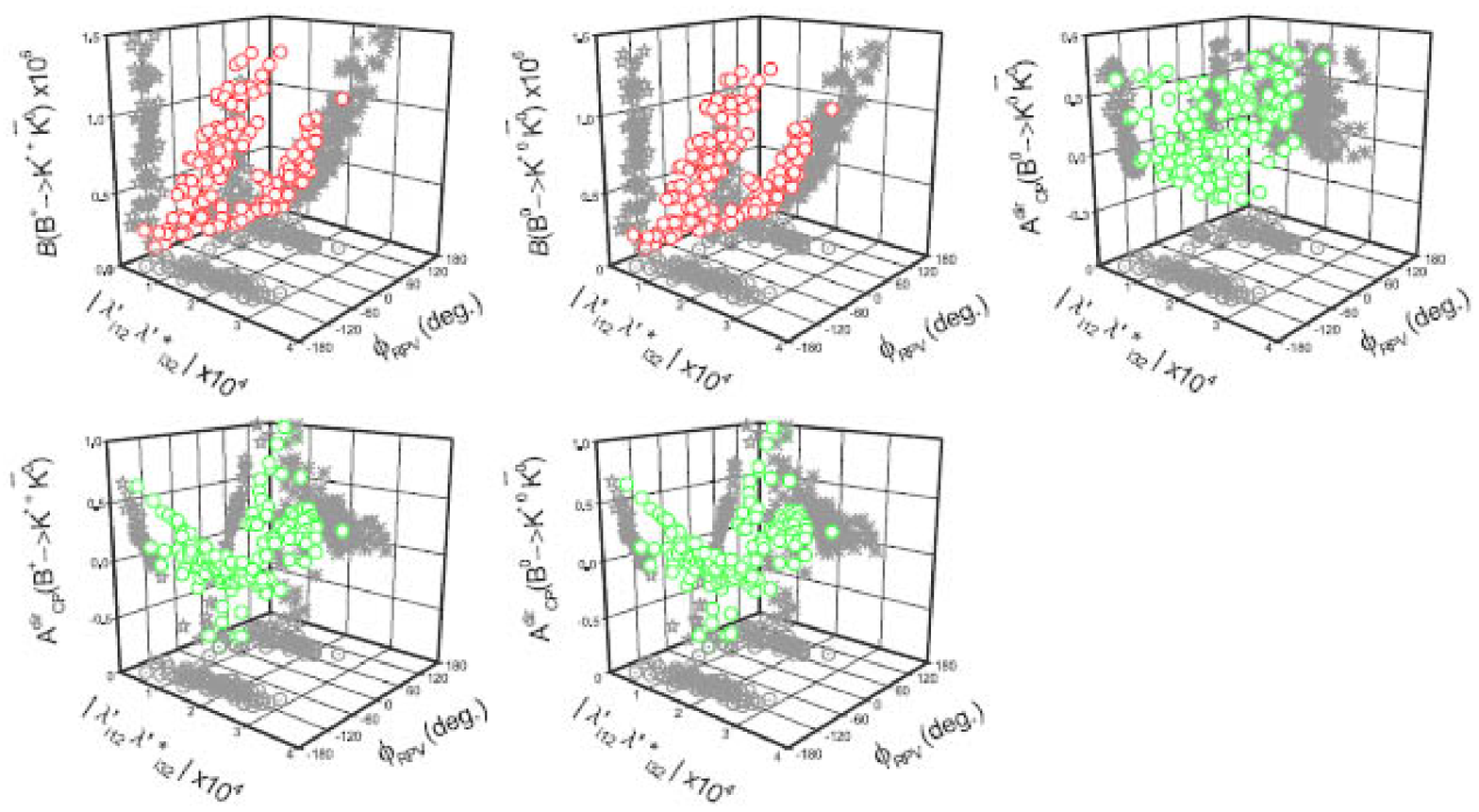}
\end{center}
\vspace{-0.6cm}
 \caption{\small The effects of RPV coupling $\lambda'_{i12}
 \lambda'^*_{i32}$ in $B\to K^{(*)}\bar{K}^{(*)}$ decays.}
 \label{li12pli32p}
\end{figure}

\begin{figure}[htbp]
\begin{center}
\includegraphics[scale=0.75]{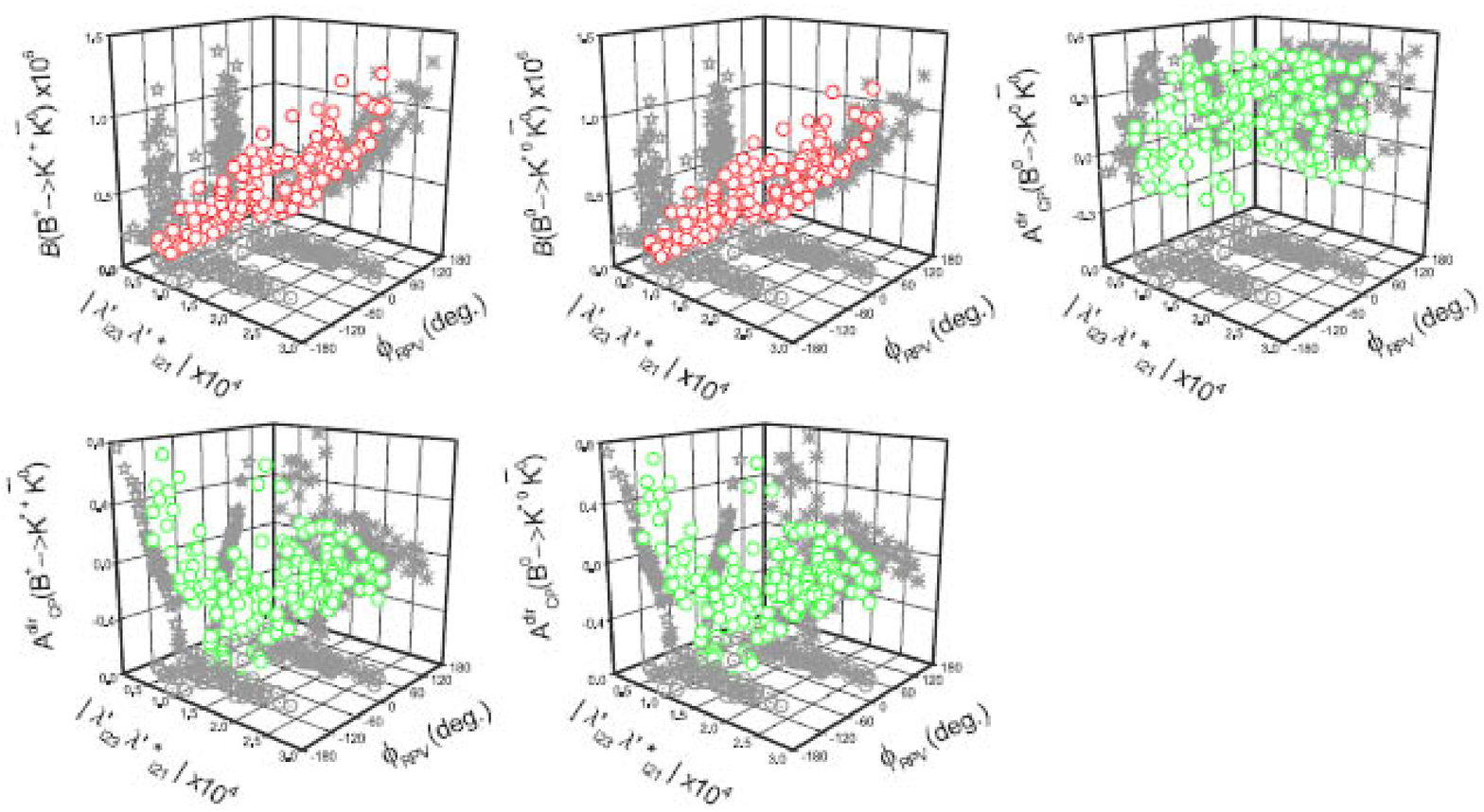}
\end{center}
\vspace{-0.6cm}
 \caption{\small The effects of RPV coupling $\lambda'_{i23}
 \lambda'^*_{i21}$ in $B\to K^{(*)}\bar{K}^{(*)}$ decays.}
 \label{li23pli21p}
\end{figure}

\end{document}